\documentclass[a4paper,11pt]{article}
\usepackage{jheppub} 

\usepackage{amsfonts}
\usepackage{amsmath}
\allowdisplaybreaks[4]        
\usepackage{amssymb}
\usepackage{euscript}     
\usepackage{xcolor}         
\usepackage{tensor}        
\usepackage{amsthm}
\usepackage{graphicx}
\usepackage{subfigure}
\usepackage{bbm}
\usepackage[header,title,page,titletoc]{appendix}  

\usepackage[numbers]{natbib}  
\usepackage{float}

\usepackage{hyperref}

\newcommand{\be}{\begin{equation}}
\newcommand{\bea}{\begin{eqnarray}}
\newcommand{\eea}{\end{eqnarray}}
\newcommand{\ba}{\begin{array}}
\newcommand{\ea}{\end{array}}
\newcommand{\ee}{\end{equation}}
\newcommand{\bes}{\begin{equation*}}
\newcommand{\beas}{\begin{eqnarray*}}
\newcommand{\eeas}{\end{eqnarray*}}
\newcommand{\bas}{\begin{array*}}
\newcommand{\eas}{\end{array*}}
\newcommand{\ees}{\end{equation*}}

\title{\boldmath Entropy of Hawking Radiation for Two-Sided Hyperscaling Violating Black Branes}

\author[]{Farzad Omidi}

\affiliation[]{School of Physics, Institute for Research in Fundamental Sciences (IPM),\\
	P.O. Box 19395-5531, Tehran, Iran}

\emailAdd{farzad@ipm.ir}

\abstract{
In this paper, we study the von Neumann entropy of Hawking radiation $S_{\rm R}$ for a $d+2$-dimensional Hyperscaling Violating (HV) black brane which is coupled to two Minkowski spacetimes as the thermal baths. We consider two different situations for the matter fields: First, they are described by a $CFT_{d+2}$ whose central charge $c$ is very large. Second, they are described by a d+2 dimensional HV QFT which has a holographic gravitational theory that is a HV geometry at zero temperature. For both cases, we calculate the Page curve of the Hawking radiation as well as the Page time $t_{\rm Page}$. For the first case, $S_{\rm R}$ grows linearly with time before the Page time and saturates after this time. Moreover, $t_{\rm Page}$ is proportional to $\frac{2 S_{\rm th}}{c T}$, where $S_{\rm th}$ and $T$ are the thermal entropy and temperature of the black brane. For the second case, when the hyperscaling violation exponent $\theta_m$ of the matter fields is zero, the results are very similar to those for the first case. However, when $\theta_m \neq 0$, the entropy of Hawking radiation grows exponentially before $t_{\rm Page}$ and saturates after this time. Furthermore, the Page time is proportional to $\log \left( \frac{1}{G_{\rm N,r}} \right) $, where $G_{\rm N,r}$ is the renormalized Newton's constant. It was also observed that for both cases, $t_{\rm Page}$ is a decreasing and an increasing function of the dynamical exponent $z$ and hyperscaling violation exponent $\theta$ of the black brane geometry, respectively. Moreover, for the second case, $t_{\rm Page}$ is independent of $z_m$, and for $\theta_m \neq 0$, it is a decreasing function of $\theta_m$.}


\arxivnumber{2112.05890}

\begin{document} 

   \begin{flushright}
   IPM/P-2021/41 \\	
   \end{flushright}

   \maketitle
    \flushbottom


\section{Introduction}
\label{Sec: Introduction}

It is well known that black holes can emit particles with a thermal spectrum which is called Hawking radiation \cite{Hawking:1975vcx}. In this manner, a black hole loses its mass, and might be disappeared completely. Assuming that the black hole was formed from the collapse of some matter in a pure quantum state leads to the famous Hawking's information paradox \cite{Hawking:1976ra}, since the final state, i.e. the emitted radiation, is in a mixed state. Moreover, the final state cannot be obtained via a S matrix from the initial pure state. On the other hand, the fine-grained (von Neumann) entropy of the Hawking radiation $S_{\rm R}$ grows linearly in time until the black hole completely disappears.  In this manner, it eventually exceeds the coarse-grained (thermodynamic or Bekenstein-Hawking) entropy $S_{\rm BH}$ of the initial black hole. On the other hand, the von Neumann entropies of the black hole and radiation are equal to each other, i.e. $S_{\rm R} = S_{\rm black \; hole}$, since the whole system is in a pure state. Moreover, the fine-grained entropy of the black hole has to be less than its coarse-grained entropy, i.e. $S_{\rm black \; hole} \leq S_{\rm BH}$ \cite{Page:2013dx}. Therefore, one has \cite{Page:2013dx} (See also \cite{Almheiri:2020cfm})
\bea
S_{\rm R} \leq S_{\rm BH}.
\label{S-R leq S-BH-one sided}
\eea
\footnote{Notice that this equation is only valid for one-sided black holes. For two-sided black holes, it has to be modified to  \cite{Almheiri:2019yqk}
	\bea
	S_{\rm R} \leq 2 S_{\rm BH}.
	\label{S-R leq-SBH-two sided}
	\eea 
	In this case, $S_{\rm R}$ saturates at $2  S_{\rm BH}$ after the Page time and it does not decrease in contrast to that for one-sided black holes.}
The time at which the saturation happens is called the "Page time" \cite{Page:1993wv,Page:2013dx}. Moreover, since the $S_{\rm BH}$ is a decreasing function of time, $S_{\rm R}$ has to start decreasing after the Page time and goes to zero at the end of the evaporation. Consequently, the unitarity requires that the fine-grained entropy of Hawking radiation follows the so called Page curve \cite{Page:1993wv,Page:2013dx}.
\\ Unearthing the unitary evaporation of Black holes is one of the most important puzzles in gravity. Recently, a very sophisticated resolution for information paradox which is called "{\it island rule}" were introduced \cite{Penington:2019npb,Almheiri:2019psf,Almheiri:2019hni,Almheiri:2019yqk} in the context of the AdS/CFT correspondence \cite{Maldacena:1997re}. This rule is inspired by the concept of "Quantum Extremal Surfaces" (QES) \cite{Engelhardt:2014gca} which are the generalization of the (H)RT surfaces \cite{Ryu:2006bv,Hubeny:2007xt} applied in the calculation of holographic entanglement entropy. More precisely, a QES is a classical spacelike codimension-two surface in the bulk spacetime which minimizes the generalized entropy \cite{Engelhardt:2014gca,Faulkner:2013ana} which is composed of an area term and the von Neumann entropy of matter quantum fields in a bulk region enclosed between the QES and the corresponding boundary region. The motivation for the island rule was the observation that the entanglement wedge (EW) of an evaporating black hole at late times does not include all of its interior. Therefore, one might naturally expects that the EW of the radiation region $\mathcal{R}$, which is a spatial region far away from the black hole where the radiation is collected, contains some parts of the black hole interior dubbed "island" $\mathcal{I}$ \cite{Penington:2019npb, Almheiri:2019hni} (See figure \ref{fig: Penrose}). Consequently, to calculate the von Neumann entropy of the Hawking radiation, one has to consider the contributions of islands. 
According to the rule, one should first calculate the generalized entropy $S_{\rm gen}$
as follows \cite{Almheiri:2019hni,Almheiri:2019yqk}
\bea
S_{\rm gen} 
\left( \mathcal{R} \cup \mathcal{I} \right)
 = \frac{Area(\partial I)}{4 G_N} + S_{\rm matter} \left( \mathcal{R} \cup \mathcal{I} \right),
\label{S-gen}
\eea 
where $\partial \mathcal{I}$ is the boundary of the island and is a codimension-two surface. It is expected that for one-sided black holes $\partial \mathcal{I}$ is located inside the black hole \cite{Penington:2019npb,Almheiri:2019psf,Almheiri:2019hni,Gautason:2020tmk,Hartman:2020swn}. However, for two-sided black holes, it is located outside the black hole \cite{Almheiri:2019yqk,Almheiri:2019psy,Gautason:2020tmk}. If the matter quantum field theory has a gravitational dual theory, the island is connected to the radiation region through extra dimensions \cite{Almheiri:2019hni}. Moreover, $S_{\rm matter} \left( \mathcal{R} \cup \mathcal{I} \right)$ is the von Neumann entropy of matter quantum fields in the region $\mathcal{R} \cup \mathcal{I}$. In the following, we denote the endpoints of $\mathcal{R}$ and $\mathcal{I}$ by $b_{\pm}$ and $a_{\pm }$, respectively (See figure \ref{fig: Penrose}). Next, one has to extremize $S_{\rm gen}$ over all possible islands. If there is more than an island, one should consider the one which gives the minimum generalized entropy
\bea
S_{\rm R}=  {\rm min}_{\mathcal{I}} \lbrace {\rm ext}_{\mathcal{I}} \left[ S_{ \rm gen} \right] \rbrace.
\label{S-R}
\eea 
In this case, $\partial \mathcal{I}$ is a minimal QES. Moreover, when there are no islands, the area term in eq. \eqref{S-gen} vanishes, and hence the entropy of Hawking radiation is simply given by the von Neumann entropy of the matter fields on the region $\mathcal{R}$
\bea
S_{\rm R}= S_{\rm matter} \left( \mathcal{R} \right).
\label{S-R-no island}
\eea 
The island rule were verified for two-dimensional Jackiw-Teitelboim (JT) black holes \cite{Jackiw:1984je,Teitelboim:1983ux} in refs. \cite{Penington:2019kki,Almheiri:2019qdq,Goto:2020wnk}
\footnote{See also \cite{Marolf:2020rpm,Bousso:2021sji} for related discussions. Moreover, a new proof was recently introduced in ref. \cite{Pedraza:2021ssc} which is based on minimizing the microcanonical action of an entanglement wedge.}
 by calculating the Euclidean gravitational path integral via replica trick in gravity \cite{Lewkowycz:2013nqa,Dong:2016hjy}. It was observed that the gravitational path integral has two saddle points: First, Hawking saddle which gives the usual entropy for Hawking radiation that grows linearly with time. Second, replica wormhole saddle which is a wormhole connecting various copies of the original black hole. It was observed that the replica saddle leads to the contribution of the islands and enforces the entropy of radiation to saturate. It should be pointed out that the Hawking saddle is dominant before the Page time. However, the replica wormhole saddle is dominant after the Page time. 
\\Furthermore, the island rule have been explored extensively for various black holes in flat and AdS spacetimes such as:
JT gravity \cite{Penington:2019kki,Almheiri:2019qdq,Almheiri:2019yqk,Hollowood:2020cou,Chen:2020jvn,Goto:2020wnk,Chen:2019uhq,Balasubramanian:2021xcm}, two-dimensional dilaton gravity
\cite{Gautason:2020tmk,Anegawa:2020ezn,Hartman:2020swn,Wang:2021mqq,He:2021mst}
, in higher dimensions \cite{Almheiri:2019psy,Hashimoto:2020cas,Karananas:2020fwx,He:2021mst,Chen:2020uac,Chen:2020hmv,Krishnan:2020fer,Geng:2020qvw,Matsuo:2020ypv,Ghosh:2021axl,Arefeva:2021kfx,Saha:2021ohr}, higher derivative gravity theories \cite{Alishahiha:2020qza}, charged black holes \cite{Ling:2020laa,Wang:2021woy,Kim:2021gzd,Ahn:2021chg,Yu:2021cgi}, pure BTZ black hole microstates
\footnote{These geometries are obtained by excising a two-sided BTZ black hole with a timelike dynamical brane, dubbed End-of-the-World (ETW) brane \cite{Kourkoulou:2017zaj,Almheiri:2018ijj,Cooper:2018cmb}. The action-complexity of this model was studied in ref \cite{Omidi:2020oit}.}
\cite{Balasubramanian:2020hfs,Anderson:2020vwi,Fallows:2021sge}, and black holes coupled to gravitating baths \cite{Balasubramanian:2021wgd,Anderson:2020vwi,Geng:2020fxl,Geng:2021iyq}. See also \cite{Almheiri:2020cfm} for a very nice review on the topic. 
Furthermore, the rule were applied
in de Sitter spacetime \cite{Hartman:2020khs,Balasubramanian:2020xqf,Aalsma:2021bit,Sybesma:2020fxg,Geng:2021wcq,Chen:2020tes}, flat-space cosmology
\footnote{It is a solution of the three dimensional Einstein gravity with no cosmological constant.}
\cite{Azarnia:2021uch} and separate universes \cite{Balasubramanian:2021wgd,Balasubramanian:2020coy,Miyata:2021ncm,Fallows:2021sge,Miyata:2021qsm}.
\\In this paper, we study the von Neumann entropy of the Hawking radiation for a two-sided Hyperscaling Violating (HV) black brane which is coupled to two thermal baths that are Minkowski spacetimes. We assume that the matter fields in the black brane geometry as well as in the baths are described by the same QFT. We consider two different scenarios for the QFT: First, it is a d+2 dimensional CFT which does not necessarily have a dual gravity theory.
Second, it is a d+2 dimensional QFT which has a dual gravity theory that is a zero-temperature HV geometry (See eq. \eqref{metric-HV-zero-temp}). We dub it "HV QFT". We calculate $S_{\rm R}$ for the two cases and verify that it obeys eq. \eqref{S-R leq-SBH-two sided}. It is observed that at early times there are no islands. However, at late times there is an island which leads to the saturation of $S_{\rm R}$ at twice the Bekenstein-Hawking entropy of the black brane. Moreover, we study the Page curve and Page time for different values of the dynamical exponent $z$ and hyperscaling violation exponent $\theta$ of the black brane. For the case where the matter is described by a HV QFT, we also study the behavior of the Page time and Page curve with respect to the dynamical and hyperscaling violation exponents of the HV QFT which are shown by $z_m$ and $\theta_m $, respectively. We study the two cases  $\theta_m =0$ and $\theta_m \neq 0$, separately. 
\\The organization of the paper is as follows: in Section \ref{Sec: HV Black Branes}, we briefly review the HV black brane geometry. In Section \ref{Sec: Entropy of Hawking Radiation with Matter CFT}, we calculate the entropy of Hawking radiation for the case where the matter fields
are described by a $CFT_{d+2}$. We also find the Page time and study its behavior as a function of the dynamical exponent $z$ and hyperscaling violation exponent $\theta$ of the black brane geometry. In Section \ref{Sec: Entropy of Hawking Radiation with Matter HV QFT}, we do the same calculations for the case where the matter fields are described by a d+2 dimensional HV QFT.
In section \ref{Sec: Discussion}, we summarize our results and briefly address a few interesting future directions.



\section{HV Black Branes}
\label{Sec: HV Black Branes}

In this section, we briefly review Hyperscaling Violating (HV) black branes. These are solutions to the Einstein-Maxwell-Dilaton gravity with the following action \cite{Alishahiha:2012qu}
\bea
I_{\rm HV}= - \frac{1}{16 \pi G_N} \int d^{d+2} x \sqrt{-g} \left[ R - \frac{1}{2} \left( \partial \phi \right)^2 + V(\phi) - \frac{1}{4} e^{\lambda \phi} F^2 \right],
\label{action-HV}
\eea 
The gauge field $A$ breaks Lorentz invariance and introduces the dynamical exponent $z$. 
\footnote{It should be pointed out that the action is Lorentz invariant and the solution does not respect the Lorentz symmetry. More precisely, under the coordinate transformation \cite{Dong:2012se}
\bea
t \rightarrow \lambda^z t \;\;\;\;\;\;\;\;\;\;\;\;\;\;\; x^i \rightarrow \lambda x^i \;\;\;\;\;\;\;\;\;\;\;\;\;\;\; r \rightarrow \lambda r 
\nonumber
\label{coord-transformation}
\eea 	
the metric \eqref{metric-BB} is covariant, i.e. $ ds \rightarrow \lambda^{\frac{\theta}{d}} ds$. It is clear that there is an anisotropy among the $t$ and $x^i$ coordinates in the above transformation. Therefore, the Lorentz symmetry is broken when $z \neq 1$, and it is a consequence of having a non-zero gauge field in eq. \eqref{dilaton-gauge-fields} for $z \neq 1$.}
Moreover, the non-trivial potential $V(\phi)$ breaks the scaling symmetry and introduces the Hyperscaling Violation exponent $\theta$. The dilaton field $\phi$, its potential $V( \phi)$ and the field strength of the gauge field are given by (See ref. \cite{Alishahiha:2012qu} for more details)
\bea
\phi &=&  \phi_0 + \beta \ln r, \;\;\;\;\;\;\;\;\;\;\;\;\; V(\phi) =  (d_e + z -2)(d_e +z-1) e^{\gamma ( \phi - \phi_0) },
\cr && \cr
F_{ rt}  &=&  \sqrt{2(z-1) (d_e + z-1)} e^{\frac{(d_e + \theta_e-1)  \phi_0}{ \beta }} \; r^{d_e +z-2} ,
\label{dilaton-gauge-fields}
\eea 
where the constant parameters are defined as follows
\bea
\beta = \sqrt{2 (d_e-1) (-\theta_e + z -1)}, \;\;\;\;\;\;\;\;\;\;\;\;\; 
\lambda = - \frac{2 ( d_e + \theta_e-1) }{\beta}, 
\;\;\;\;\;\;\;\;\;\;\;\;\;
\gamma= \frac{2 \theta_e}{\beta}.
\eea 
On the other hand, the metric of the black brane is given by \cite{Alishahiha:2012qu}
\footnote{We set the AdS radius to one, i.e. $R=1$. Moreover, we also set the dynamical scale $r_F$ to one.}
\bea
ds^2 =  r^{- 2 \theta_e} \left( - r^{2z} f(r) dt^2  + \frac{dr^2}{r^2 f(r)} + r^2 \sum_{i=1}^{d} dx_i^2 \right),
\label{metric-BB}
\eea
where the emblankening factor is as follows
\bea
f(r) = 1- \left( \frac{r_h}{r} \right)^{d_e +z}.
\label{f(r)}
\eea 
Here we defined $\theta_e = \frac{\theta}{d}$ and an effective dimension $d_e = d- \theta$. Furthermore, the null energy condition puts the following constraints on the values of $z$ and $\theta$
\bea
d_e ( d (z-1) - \theta) \geq 0, \;\;\;\;\;\;\;\;\;\;\;\;\;\;\;\;\; (z-1)(d_e +z) \geq 0.
\label{NEC}
\eea 
In the following, we restrict ourselves to the case $z \geq 1$ and $d_e > 0$. 
\footnote{As mentioned in ref. \cite{Alishahiha:2012qu}, this solution is not valid for the case $d_e=0$. Moreover, for $d> \theta$ the geometry is unstable \cite{Dong:2012se}.}
It should also be pointed out that for $\theta=0$ and $z=1$, the scaling and Lorentz symmetries are restored in the dual HV QFT. In this case, the scalar field becomes a constant and the gauge field equals to zero. Moreover, $V(\phi)$ plays the rule of the cosmological constant $\Lambda$, and hence the solution reduces to a d+2 dimensional AdS black brane. On the other hand, the temperature and thermal (Bekenstein-Hawking) entropy of the black brane are given by \cite{Alishahiha:2012qu}
\bea
T = \frac{( d_e +z )  r_h^z}{4 \pi}, \;\;\;\;\;\;\;\;\;\;\;\;\;\;\; S_{th} = \frac{V_d r_h^{d_e}}{4 G_N}.
\label{temp}
\eea 
Here $V_d$ is the volume in the transverse directions $x^i$. On the other hand, the tortoise coordinate is as follows
\bea
r^\ast (r)= \int \frac{dr}{r^{z+1} f(r)} = {_2}F_1 \left[ 1, \frac{z}{ d_e+z} , 1+ \frac{z}{ d_e+z}, \left( \frac{r_h}{r} \right)^{d_e +z }\right].
\label{tortoise}
\eea 

\section{Entropy of Hawking Radiation: Matter CFT}
\label{Sec: Entropy of Hawking Radiation with Matter CFT}

\begin{figure}
	\begin{center}
		\includegraphics[scale=0.65]{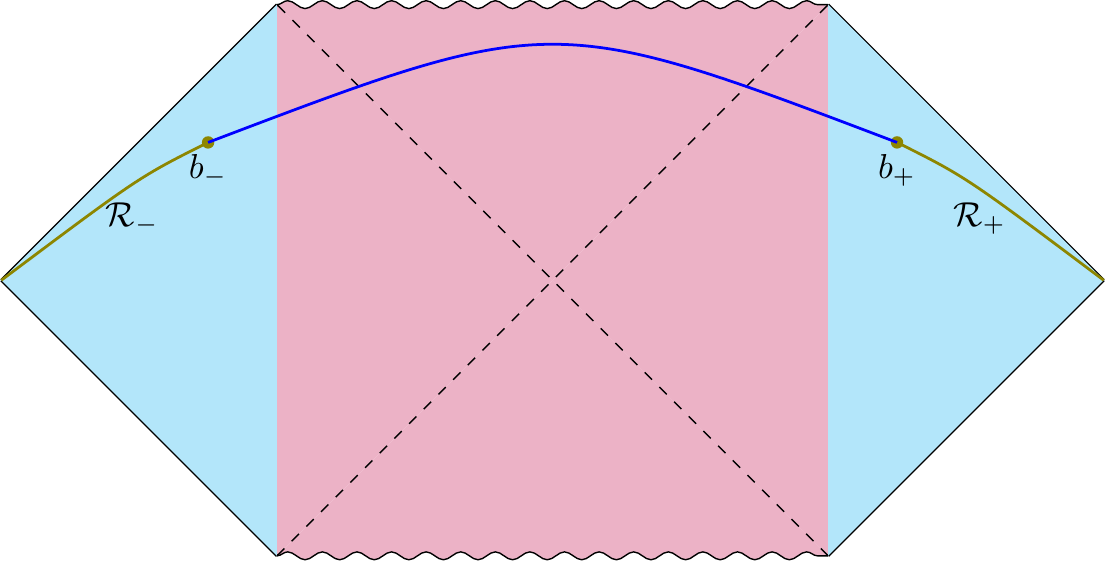}
		\hspace{0.2cm}
		\includegraphics[scale=0.65]{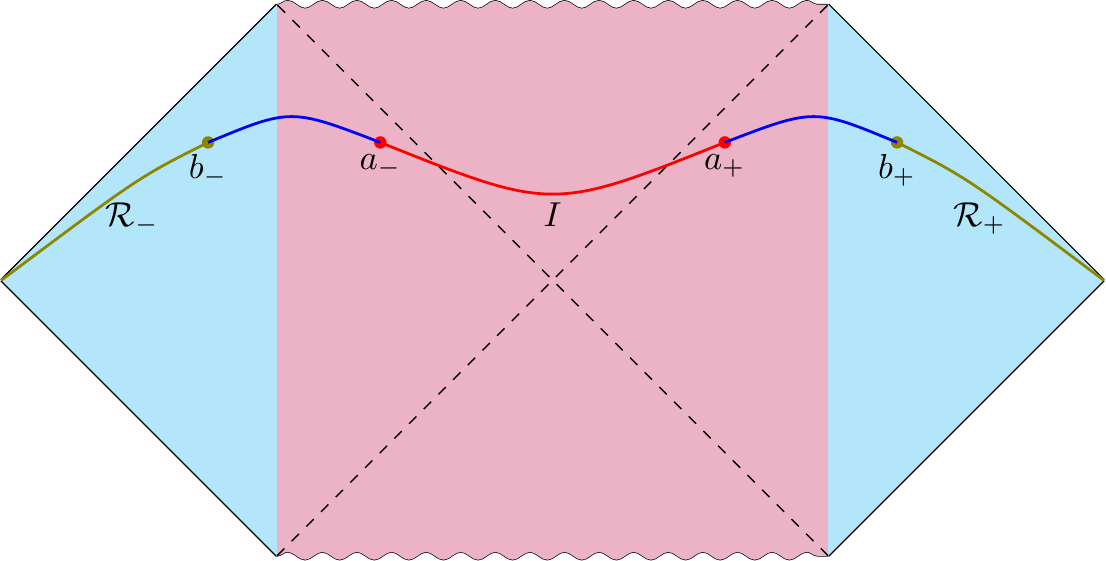}
	\end{center}
	\caption{
		The Penrose diagram of the HV black brane geometry which is shown in purple and two Minkowski spacetimes as the baths shown in cyan. {\it Left}) when there are no islands. {\it Right}) when there is an island $\mathcal{I}$ indicated in red with endpoints $a_-$ and $a_+$. The radiation region is $\mathcal{R} = \mathcal{R}_+ \cup \mathcal{R}_-$ and indicated in olive. The endpoints of $\mathcal{R}$ are denoted by $b_{\pm}$ and are located in the baths. We assume that the state on the full Cauchy slice is pure. Therefore, one can calculate the EE of matter fields $S_{\rm matter} (\mathcal{R} \cup \mathcal{I})$ on the complement intervals $[b_-,b_+]$ in the left panel and $[b_-,a_-] \cup [a_+,b_+]$ in the right panel which are shown in blue.}
	\label{fig: Penrose}
\end{figure}

In this section, we study the entropy of Hawking radiation for a two-sided HV black brane geometry. To allow the black brane to evaporate, we couple it to two thermal baths which are two Minkowski spacetimes on the left and right hand side of the Penrose diagram (See figure \ref{fig: Penrose}). Similar to ref. \cite{Almheiri:2019hni} we add some matter fields in the bulk spacetime. For convenience, we assume that the matter fields in the HV black brane geometry and inside the two baths are described by the same QFT. 
Moreover, one can impose transparent boundary conditions \cite{Almheiri:2019qdq} for the matter fields on the boundary of the HV black brane geometry. Furthermore, we assume that the matter fields are described by a $\rm CFT_{d+2}$. Then the whole action in the gravity region (see the purple region in figure \ref{fig: Penrose}) is given by
\bea
I = I_{\rm HV} + I^{\rm matter}_{CFT_{d+2}}.
\label{action}
\eea 
The CFT may \cite{Almheiri:2019hni,Almheiri:2019psy,Gautason:2020tmk,Chen:2019uhq,Chen:2020uac,Chen:2020jvn,Chen:2020hmv,Ling:2020laa} or may not \cite{Almheiri:2019yqk,Alishahiha:2020qza,Hashimoto:2020cas} have a dual gravity theory. Moreover, we assume that its central charge is very large, i.e. $c \gg 1$. Therefore, the contributions of the matter fields of the $CFT_{d+2}$ to the EE are dominant over those of the graviton and the matter fields in $I_{\rm HV}$ \cite{Almheiri:2019hni,Alishahiha:2020qza,Azarnia:2021uch}. Moreover, we assume that 
\bea
c \ll \frac{V_d r_h^{d_e} }{G_N} \propto S_{th},
\label{c ll S-th}
\eea 
to be able to neglect the backreaction of the matter on the HV black brane geometry \cite{Alishahiha:2020qza,Azarnia:2021uch}. 
\\ Furthermore, it is believed that the generalized entropy is finite and independent of the UV cutoff of the theory \cite{Bousso:2015mna,Susskind:1994sm,Engelhardt:2014gca,Almheiri:2020cfm}. On the other hand, there are some UV divergent terms in the entanglement entropy (EE) of a QFT, and the leading divergent term follows the area law \cite{Bombelli:1986rw,Srednicki:1993im}. Notice that our entangling regions are in the shape of strips on a constant time slice
such that they are extended in the transverse directions $x^i$ (See figure \ref{fig: Penrose}). Moreover, from holographic calculations, it is known that the EE of a strip with width $\ell$ and lengths $L$ in a holographic $CFT_{d+2}$ with $d \geq 1$ which is in its vacuum state, is given by
\cite{Ryu:2006ef}
\footnote{For spherical entangling regions there are subleading UV divergent terms \cite{Ryu:2006ef}.}
\bea
S= \frac{1}{2 G_N^{d+3} } \Bigg[ \frac{2 R^{d+1}}{d} \left( \frac{L}{ \epsilon} \right)^d - \frac{(2 \sqrt{ \pi }R)^{d+1}}{d} \left( \frac{ \Gamma \left( \frac{d+2}{2 (d+1)}\right)}{ \Gamma \left( \frac{1}{2 (d+1)}\right)}\right) \left( \frac{L}{\ell} \right)^d
\label{HEE-strip}
\Bigg].
\eea 
Here the first term is proportional to the area $L^{d}$ of the strip and the second term is a constant universal term. Inspired by this observation, one might expect that 
\bea
S_{\rm matter} ( \mathcal{R} \cup \mathcal{I}) = \frac{Area ( \partial \mathcal{I})}{ \epsilon^d} + S_{\rm matter}^{\rm f} ( \mathcal{R} \cup \mathcal{I}),
\label{S-matter}
\eea 
where $\epsilon$ is the UV cutoff of the matter $CFT_{d+2}$.
Therefore, we renormalize Newton's constant as follows \cite{Susskind:1994sm,Almheiri:2020cfm,Hashimoto:2020cas,Alishahiha:2020qza,Azarnia:2021uch}
\footnote{It was shown in ref. \cite{Susskind:1994sm} that in four dimensions, i.e. for d=2, the loop diagrams for matter fields lead to the renormalization of the gravitational coupling $G_N$ as in eq. \eqref{G_N-r-CFT}. Moreover, this renormalization cancel the area law divergent term in eq. \eqref{S-matter} such that the generalized entropy is finite. See \cite{Bousso:2015mna} for a nice review on this topic.}
\bea
\frac{1}{4 G_{ N,r}} = \frac{1}{4 G_N} - \frac{1}{\epsilon^{d} },
\label{G_N-r-CFT}
\eea 
to absorb the UV divergent term in eq. \eqref{S-matter}. Consequently, we merely consider the finite part of the EE of the matter fields in the following, and for an arbitrary interval $[ m_1 , m_2]$ in the bulk spacetime, we denote it  by $S^{\rm f}_{\rm matter} (m_1 , m_2)$. Having said this, we rewrite the generalized entropy in eq. \eqref{S-gen} as follows \cite{Hashimoto:2020cas,Alishahiha:2020qza,Azarnia:2021uch}
\bea
S_{\rm gen}
= \frac{Area(\partial I)}{4 G_{ N,r}} + S_{\rm matter}^{\rm f} \left( \mathcal{R} \cup \mathcal{I} \right).
\label{S-gen-ren}
\eea 
Another important point is that the metric in eq. \eqref{metric-BB}, has translational symmetries along the transverse directions $x^i$. Therefore, one might expect that the locations of the endpoints of the island, and hence the generalized entropy are independent of the coordinates $x_i$.
Consequently, one might expect that the problem effectively reduces to two dimensions with coordinates $(t,r)$. 
Moreover, by compactification of the $x_i$ directions, one might consider the transverse manifold $\mathbb{R}^d$ which is parametrized by $x_i$'s, as a d dimensional sphere with a very large radius. In this manner, one might assume that the s-wave approximation mentioned in refs. \cite{Hashimoto:2020cas,Alishahiha:2020qza,He:2021mst,Penington:2019npb} can be applied.
\footnote{We would like to thank Mohsen Alishahiha and Ali Naseh for their very helpful comments on this topic.}
In other words, after the expansion of the matter fields in terms of the spherical harmonics on the large sphere, one can have both massless and massive Kaluza-Klein modes. Then, each mode behaves as an independent free field in two dimensions (with coordinates $t$ and $r$) whose mass is proportional to the inverse of the radius of the sphere \cite{Penington:2019npb}. Since the radius of the sphere is very large, the masses of the Kaluza-Klein modes have to be very small. Moreover, since the endpoints $b_{\pm}$ of the radiation region are very far from the endpoints $a_{\pm}$ of the islands
\footnote{We are assuming that $a \approx r_h$ and $b \gg r_h$, where $a$ and $b$ are the radial coordinates of the endpoints $a_{\pm}$ and $b_{\pm}$, respectively.}
, we assume that only massless modes can reach the radiation region (See also \cite{Azarnia:2021uch,Penington:2019npb,Hashimoto:2020cas}). Therefore, the contribution of the massless modes to $S_{\rm matter} (\mathcal{R} \cup \mathcal{I})$ is dominant over that of the massive modes.
\footnote{We would like to thank Ali Naseh for bringing this point to our attention.}
Consequently, one may apply the formula for a two dimensional CFT in its vacuum state to calculate $S_{\rm matter} (\mathcal{R} \cup \mathcal{I})$, which 
for an interval of length $\ell$
is given by \cite{Holzhey:1994we,Calabrese:2004eu}
\bea
S= \frac{c}{3} \log \left( \frac{\ell}{\epsilon}\right),
\label{EE-CFT-interval}
\eea 
where $c$ is the central charge and $\epsilon$ is the UV cutoff.

\subsection{Kruskal Coordinates and the Distances}
\label{Sec: Kruskal Coordinates and the Distances}
For later convenience, it is better to work in the Kruskal coordinates. The advantage of working with the Kruskal coordinates is that the state on the whole Cauchy slice, i.e. the black brane plus the baths, is the vacuum state \cite{Almheiri:2019yqk,He:2021mst}, and hence one can simply apply the EE formula for the case where the matter QFT is in the vacuum state (See e.g. eq. \eqref{EE-CFT-interval}). For the black brane, they are defined as follows
\footnote{Since the endpoints of the island are located outside the black brane horizon (See figure \ref{fig: Penrose}), we do not need the coordinates in the interior regions.}
\bea
U &= &  +  e^{- \frac{2 \pi }{ \beta} \left( t -r^*(r)\right)}, \;\;\;\;\;\;\;\;\;\;\;\;\;\;\;\;\;\;\; V=  - e^{ \frac{2 \pi }{ \beta} \left( t + r^*(r)\right)}, \;\;\;\;\;\;\;\;\; \text{the left exterior}
\cr && \cr 
U  & = &  - e^{- \frac{2 \pi }{ \beta} \left( t -r^*(r)\right)}, \;\;\;\;\;\;\;\;\;\;\;\;\;\;\;\;\;\;\; V=  + e^{ \frac{2 \pi }{ \beta} \left( t + r^*(r)\right)}, \;\;\;\;\;\;\;\;\; \text{the right exterior}
\label{Kruskal-BB}
\eea 
where $\beta = \frac{1}{T}$ and $r^*(r)$ is given by eq. \eqref{metric-black brane-Kruskal}. In the Kruskal coordinates, the metric of the black brane becomes
\bea
ds^2 = - \frac{dU dV}{W(r)^2} + r^{2 (1 - \theta_e)} d \vec{x}_d^2, \;\;\;\;\;\;\;\;\;\;\;\;\; W(r) = \frac{2 \pi e^{\frac{2 \pi}{ \beta} r^*(r)}}{ \beta r^{z- \theta_e} \sqrt{f(r)}},
\label{metric-black brane-Kruskal}
\eea 
which is conformally flat in the $t$ and $r$ plane. Moreover, $W(r)$ is the warp factor. On the other hand, for each of the two thermal baths which are Minkowski spacetimes, the tortoise coordinate is simply given by
\bea
r^\ast (r)=  r.
\eea 
On the other hand, one might consider the left and right baths as the Rindler wedges of a Minkowski spacetime \cite{Almheiri:2019yqk}. In this manner, the Kruskal coordinates can be defined as follows for the two baths (See also refs. \cite{Almheiri:2019yqk,Alishahiha:2020qza,He:2021mst})
\bea
U  & = &  + e^{- \frac{2 \pi }{ \beta} \left( t - r \right)}, \;\;\;\;\;\;\;\;\;\;\;\;\;\;\;\;\;\;\; V =  - e^{ \frac{2 \pi }{ \beta} \left( t + r \right)}, \;\;\;\;\;\; \text{Left bath}
\cr && \cr
U  & = &  - e^{- \frac{2 \pi }{ \beta} \left( t - r \right)}, \;\;\;\;\;\;\;\;\;\;\;\;\;\;\;\;\;\;\; V=  + e^{ \frac{2 \pi }{ \beta} \left( t + r \right)}, \;\;\;\;\;\; \text{Right bath}
\label{Kruskal-bath}
\eea 
Thus, the metric of each bath is given by (See also \cite{Azarnia:2021uch})
\bea
ds^2 = - \frac{dU dV}{W_B (r)^2} + r^2 d \vec{x}_d^2, \;\;\;\;\;\;\;\;\;\;\;\;\; W_B(r) = \frac{2 \pi e^{\frac{2 \pi}{ \beta} r}}{ \beta}.
\label{metric-bath-Kruskal}
\eea 
To calculate the entropy of Hawking radiation, one also needs to know the distances $d(a,b)$ among the endpoints of the radiation region $\mathcal{R}$ and island $\mathcal{I}$. We indicate the endpoints of the radiation region by $b_{\pm}$ and those of the island region by $a_{\pm}$. From figure \ref{fig: Penrose}, one can easily find the $(t,r)$ coordinates of the endpoints as follows (See also \cite{Balasubramanian:2021xcm})
\bea
&&a_+ :( t_a, a) ,\;\;\;\;\;\;\;\;\;\;\;\;  a_-: ( - t_a + i \beta /2, a),
\cr && \cr 
&& b_+ :(t_b, b), \;\;\;\;\;\;\;\;\;\;\;\;  b_-: (- t_b + i \beta /2, b).
\label{a,b}
\eea 
From eqs. \eqref{metric-black brane-Kruskal} and \eqref{metric-bath-Kruskal}, one can write the distance $d(m_1,m_2)$ between two arbitrary points $m_1$ and $m_2$ in the bulk spacetime as follows
\bea
d^2 (m_1,m_2) = \frac{\left( U(m_2) - U(m_1) \right) \left( V(m_1) - V(m_2) \right)}{ W(m_2) W(m_1)}.
\label{geo-lenght}
\eea 
It is straightforward to verify that the distances are given by
\bea
\label{d}
d (a_+ ,a_-)  &=& \frac{ \beta a^{(z - \theta_e)} \sqrt{f(a)}}{ \pi} \cosh \left( \frac{2 \pi t_a}{ \beta} \right),
\\
d (b_+ ,b_-)  &=&  \frac{ \beta }{ \pi} \cosh \left( \frac{2 \pi t_b}{ \beta} \right),
\cr && \cr 
d^2(a_+ ,b_+)  &=&  d^2(a_-,b_-) =  \frac{ \beta^2 a^{(z - \theta_e)} \sqrt{f(a)}}{2 \pi^2} \left( \!\! \cosh \left( \frac{2 \pi}{ \beta} (b - r^*(a)) \right) \! - \! \cosh \left( \frac{2 \pi}{ \beta} (t_a - t_b) \right) \!\!\right) \!\! ,
\cr && \cr
d^2(a_+ ,b_-)  &=& d^2(a_-,b_+) =  \frac{ \beta^2 a^{(z - \theta_e)} \sqrt{f(a)}}{2 \pi^2} \left( \!\! \cosh \left( \frac{2 \pi}{ \beta} (b - r^*(a)) \right) \!\! + \! \cosh \left( \frac{2 \pi}{ \beta} (t_a + t_b) \right) \! \!\right) \!\! .
\nonumber
\eea 

\subsection{No Islands}
\label{Sec: Matter CFT-No Islands}

When there are no islands, there are only the radiation regions $ \mathcal{R} = \mathcal{R}_- \cup \mathcal{R}_+$ (See the left panel of figure \ref{fig: Penrose}). In this case, the area term in eq. \eqref{S-gen} vanishes, and hence
the entropy of Hawking radiation is simply given by eq. \eqref{S-R-no island}. Therefore, one has
\bea
S_{ \rm R} = S_{\rm matter}^{\rm f} \left( \mathcal{R} \right) = S_{\rm matter}^{\rm f} (b_+ ,b_-),
\label{S-R-CFT-no island-1}
\eea 
where $S(b_+ , b_-)$ is the EE of matter fields on the interval $[b_- , b_+]$. In the last equality, we applied the fact that the whole state on the Cauchy slice is pure. Therefore, one can calculate the EE of matter fields on the interval $[b_- , b_+]$ instead of the region $\mathcal{R}$ (See the left panel of figure \ref{fig: Penrose}).
Next, by applying eqs. \eqref{EE-CFT-interval} and \eqref{d}, one can rewrite eq. \eqref{S-R-CFT-no island-1} as follows
\bea
S_{ \rm R} = \frac{c}{3} \log d(b_+, b_-) 
= \frac{c}{3} \log  \left( \frac{\beta}{\pi} \cosh \left( \frac{2 \pi t_b}{\beta}\right) \right).
\label{S-R-CFT-no islands-2}
\eea 
At early times, i.e. $t_b T \ll 1$, it grows quadratically in time
\bea
S_{ \rm R} =   \frac{c}{3} \left[ \frac{(d_e+z)^2 r_h^{2z}}{8} t_b^2 + \log \left( \frac{4}{ ( d_e+z)  r_h^z }\right) \right].
\label{S-R-CFT-no islands-early-times}
\eea 
On the other hand, at late times, i.e. $t_b T \gg 1$, it grows linearly in time
\bea
S_{ \rm R}  \simeq  \frac{c (d_e+ z) r_h^z}{6} t_b.
\label{S-R-CFT-no islands-late-times}
\eea 
Therefore, the entropy of Hawking radiation exceeds the coarse-grained entropy of two black branes and information paradox is inevitable. To resolve the issue, we consider the contribution of an island to the entropy of the Hawking radiation in the next section. We will see that in this manner the entropy of Hawking radiation saturates after the Page time.

\subsection{With Islands}
\label{Sec: Matter CFT-Islands}

\begin{figure}
	\begin{center}
		\includegraphics[scale=0.37]{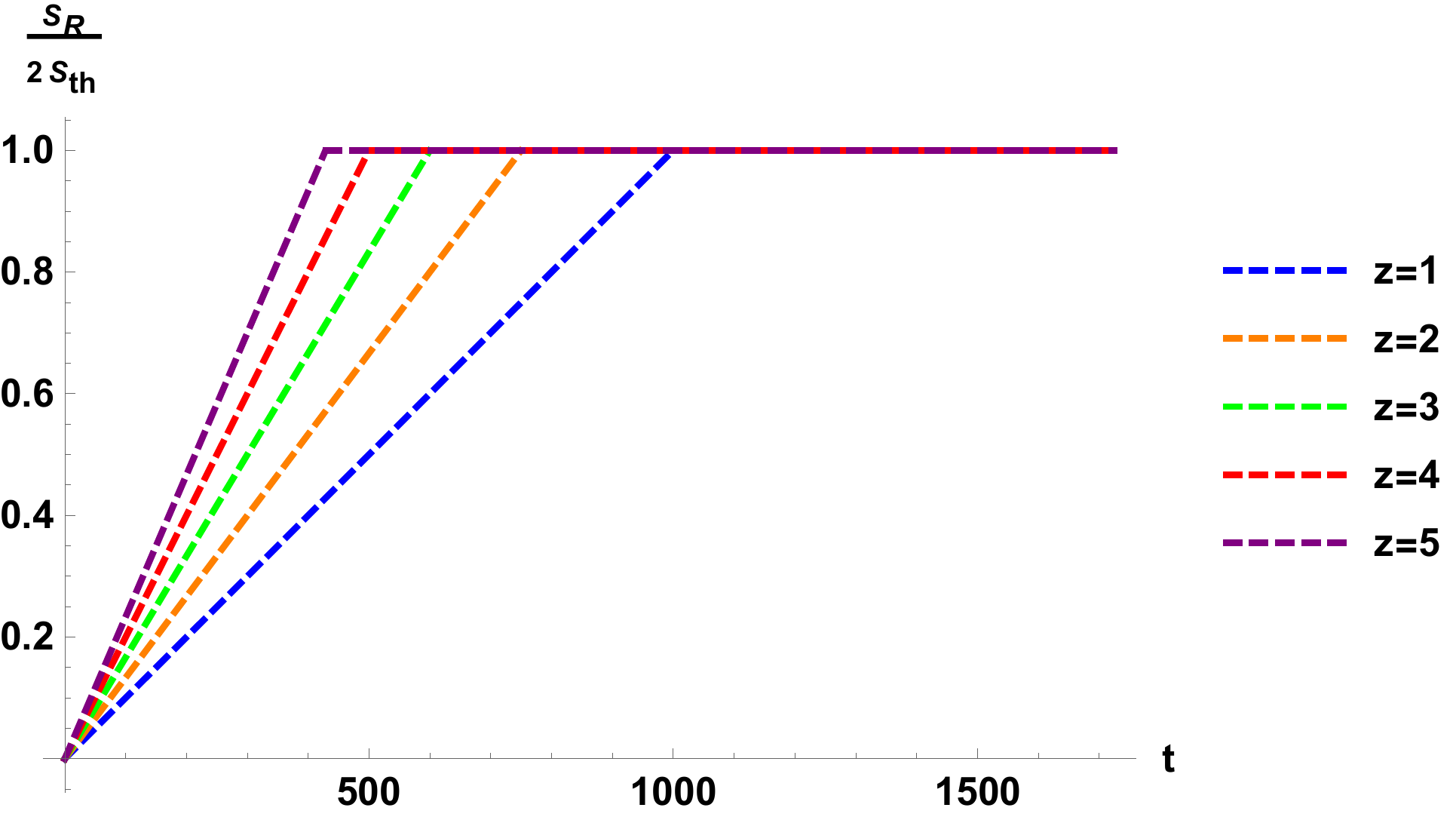}
		\includegraphics[scale=0.37]{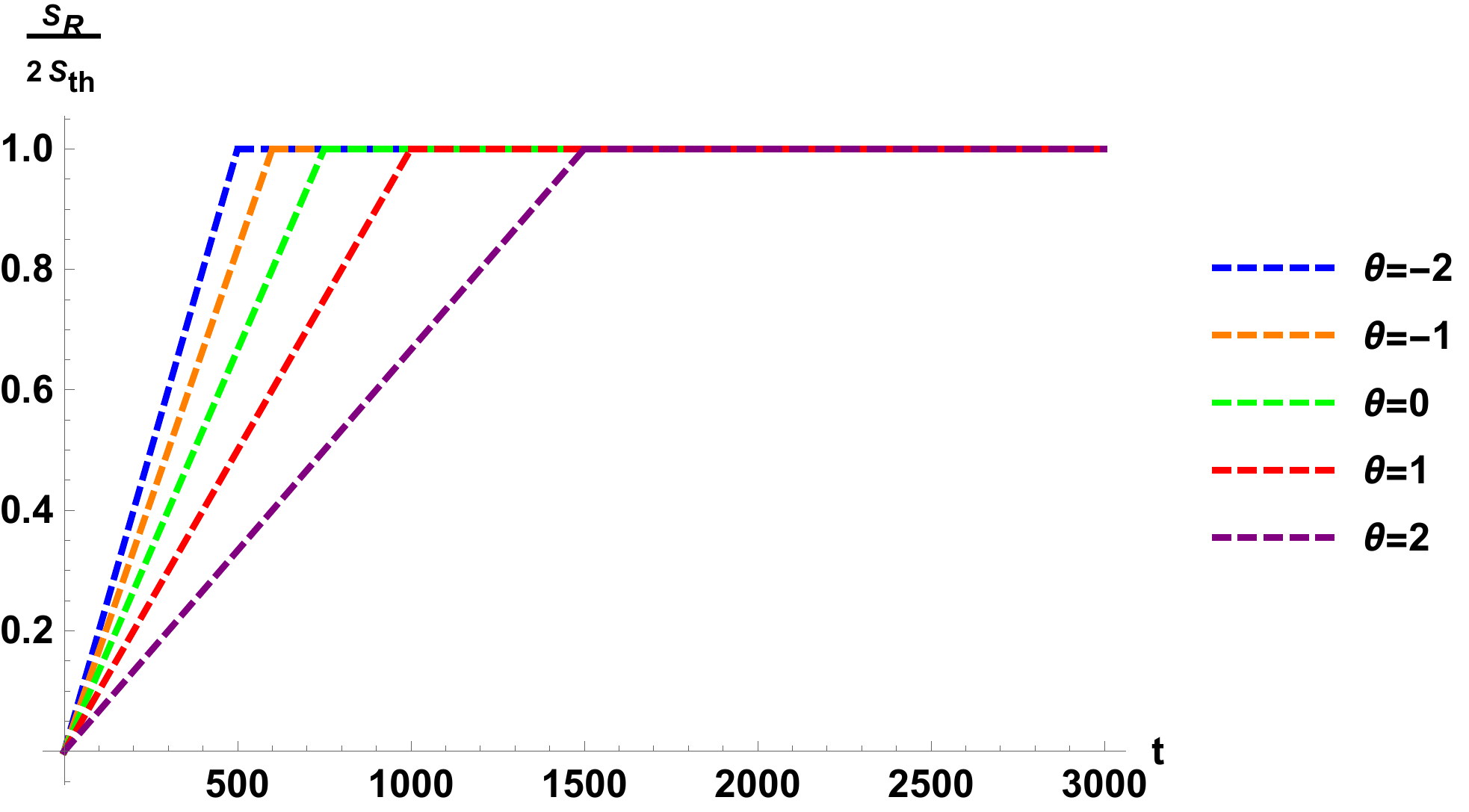}
	\end{center}
	\caption{
		Entropy of Hawking radiation $S_{\rm R}$ when the the matter is described by a $CFT_{d+2}$ or a holographic HV $QFT_{d+2}$ with $\theta_m = 0$ for: {\it Left}) $d=2$, $\theta=-1$ and different values of $z$. {\it Right}) $d=3$, $z=1$ and some values of $\theta$. Before the Page time, $S_{\rm R}$ grows linearly with time. However, it saturates to $2 S_{\rm th}$ after this time. Here we set $c=A=V_d=r_h=1$ and $G_{N,r} =0.001$. It should be pointed out we omitted the correction of order $\mathcal{O}(G_{N,r}^0)$ in eqs. \eqref{S-R-CFT-island-late-times} and \eqref{S-R-theta-0-island}, since we assumed that eq. \eqref{c ll S-th} is valid. Furthermore, we restricted ourselves to $d_e>0$ and $z \geq 1$.
	}
	\label{fig: S-R-CFT}
\end{figure}

\begin{figure}
	\begin{center}
		\includegraphics[scale=0.37]{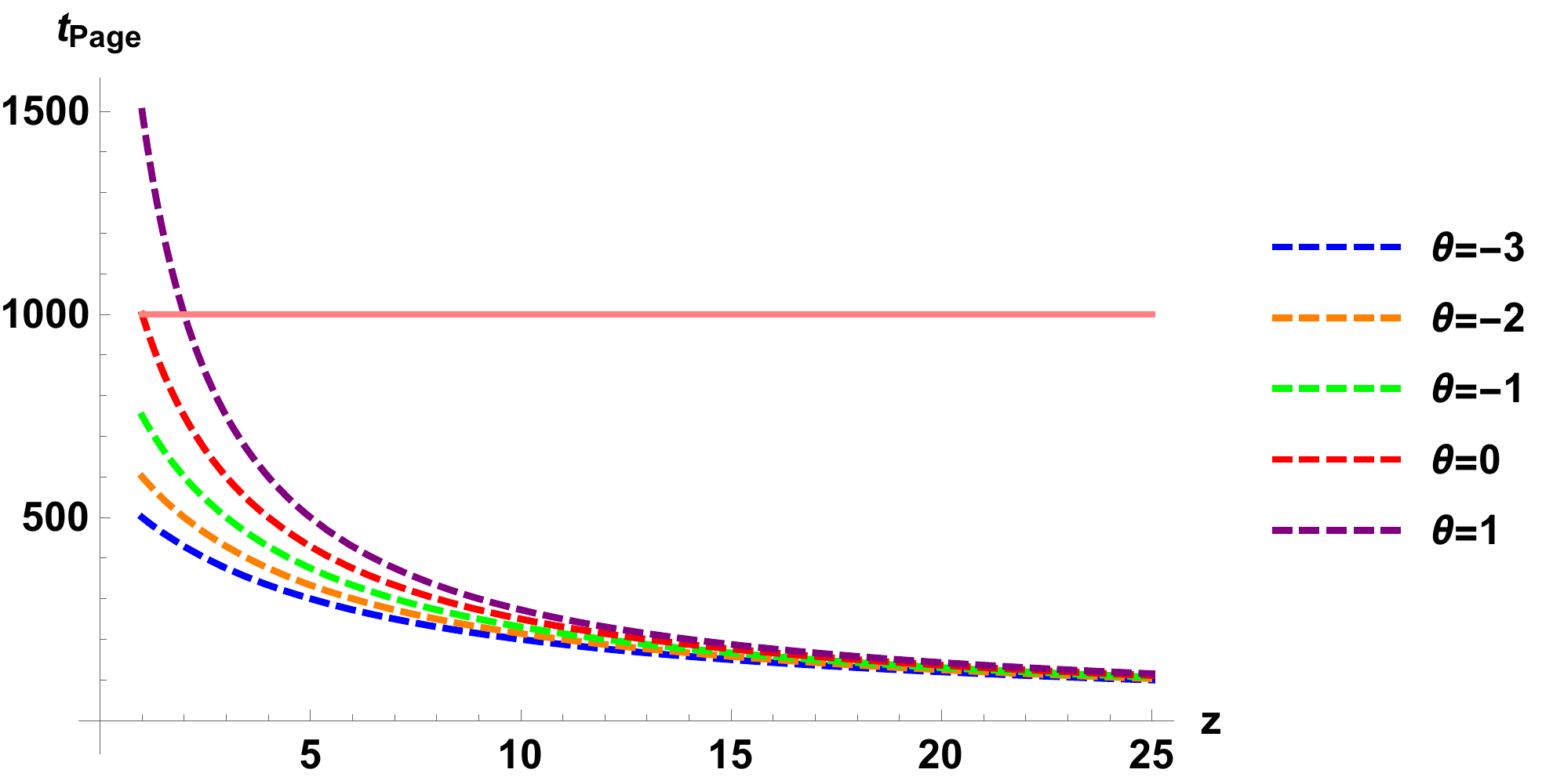}
		\includegraphics[scale=0.37]{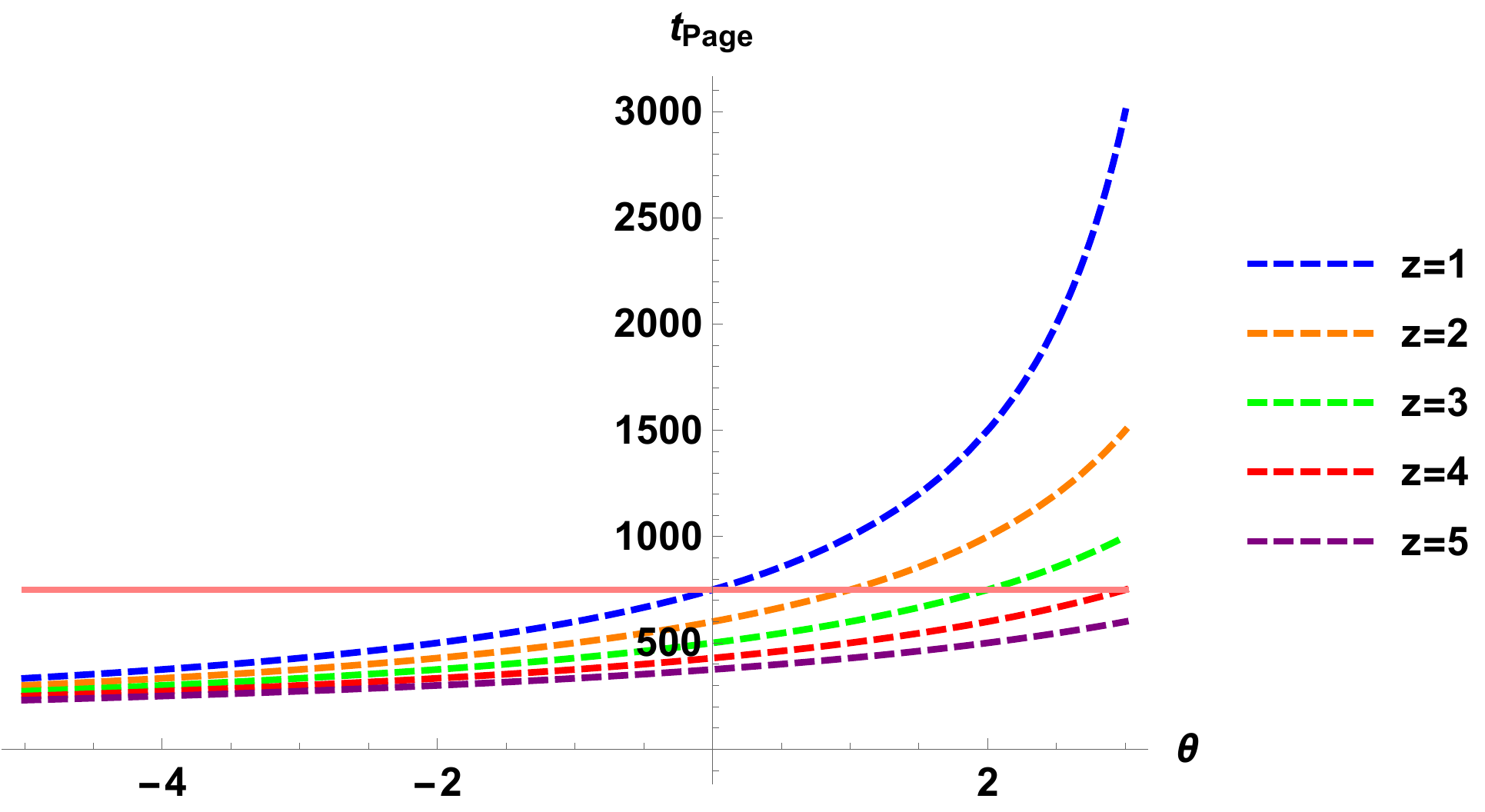}
	\end{center}
	\caption{
		Page time when the matter is described by a $CFT_{d+2}$ or a holographic HV $QFT_{d+2}$ with $\theta_m = 0$ as a function of {\it Left}) $z$ for $d=2$ and different values of $\theta$. {\it Right}) $\theta$ for $d=3$ and different values of $z$. The solid pink line indicates $t_{\rm Page}$ for the case $\theta=0$ and $z=1$. It is evident that $t_{\rm Page}$ is a decreasing and increasing function of $z$ and $ \theta$, respectively. Here we set $c=A=V_d=r_h=1$ and $G_{N,r} =0.001$. Moreover, we restricted ourselves to $d_e>0$ and $z \geq 1$.
	}
	\label{fig: t-Page-CFT}
\end{figure}

In this section, we consider the effect of islands on the entropy of Hawking radiation. We assume that there is an island $\mathcal{I}$ whose endpoints denoted by $a_-$ and $a_+$ are located outside the event horizon (See the right panel of figure \ref{fig: Penrose}). Since the state on the whole Cauchy slice is pure, one has to calculate the EE of matter fields on two disjoint intervals $[b_-,a_-] \cup [a_+ , b_+]$. 
On the other hand, as mentioned below eq. \eqref{S-gen-ren}, we are using the fact that our calculations are reduced to two dimensions with coordinates $(U,V)$. Therefore, we are effectively working with a two-dimensional matter CFT (See also \cite{Hashimoto:2020cas,Alishahiha:2020qza,Azarnia:2021uch}). As far as we are aware, the calculation of the EE for two disjoint intervals has not been fully understood yet. However, it was calculated  for free massless fermions \cite{Casini:2005rm,Casini:2009vk,Casini:2008wt} and Luttinger liquids, i.e. free compactified bosons \cite{Calabrese:2009ez} (See also \cite{Calabrese:2009qy} for a detailed review). In the following, we assume that the quantum fields of the matter CFT are $N$ free massless Dirac fermions similar to refs. \cite{Almheiri:2019qdq,Yu:2021cgi}.
In this case, for the two intervals $A = \left[ a_1, b_1 \right]$ and $B = \left[ a_2 , b_2 \right]$, one has \cite{Casini:2005rm,Casini:2009vk,Casini:2008wt}
\footnote{
	It should be pointed out that in the calculation of the EE of two disjoint intervals in the CFT of Luttinger liquids via the replica trick, there is a function $\mathcal{F}_n (x)$ in $tr (\rho_A^n) $. Here $x$ is the cross-ratio of the four endpoints of the two intervals, $n$ denotes the number of the replication and $\rho_A$ is the reduced density matrix of the two intervals.
	The function $\mathcal{F}_n (x)$ depends on the full operator content of the QFT and is unknown generally \cite{Calabrese:2009ez}. It has the property, $\mathcal{F}_n(0) =1$. Moreover, its contribution is not included in eq. \eqref{EE-CFT-two disjoint interval}, and hence, the equation has to be modified in this case (See \cite{Calabrese:2009ez,Calabrese:2009qy} for more discussions). However, at late times, i.e. $t_{a,b} T \gg 1$, the two intervals become very far apart from each other. In this case, from eqs. \eqref{d}, \eqref{approx-late-times-2-1} and \eqref{approx-late-times-2-2}, one has (See also refs. \cite{Hashimoto:2020cas,Hartman:2020swn,Almheiri:2019yqk})
	\bea
	x= \frac{d(b_-,a_-) d(b_+,a_+)}{d(b_-,a_+) d(a_-,b_+)} = \frac{d^2 (b_+, a_+)}{d^2 (a_- , b_+)} \approx \frac{ e^{\frac{2 \pi}{ \beta} (b - r^*(a))} }{\cosh \left( \frac{2 \pi }{ \beta} (t_a +t_b) \right) } \approx \frac{e^{\frac{2 \pi}{ \beta} (b - r^*(a))} }{e^{ \frac{ 4 \pi }{ \beta} t_b} } \ll 1. 
	\eea 
	Therefore,  
	$x \rightarrow 0$ and $\mathcal{F}_n(x) \rightarrow 1$. In other words, for free compactified bosons, eq. \eqref{EE-CFT-two disjoint interval} is still valid at late times \cite{Alishahiha:2020qza,Hartman:2020swn,Almheiri:2019qdq,Azarnia:2021uch}. 
We thank the referee for her/his very helpful comments on this point.}
\bea 
S (A \cup B) = \frac{c}{3} \log \left( \frac{(b_1 - a_1) (b_1 - a_2) (b_2 -a_1) (b_2 - a_2)}{\epsilon^2 (a_2 - a_1) (b_2 - b_1)} \right),
\label{EE-CFT-two disjoint interval-1}
\eea 
where $c$ is the central charge and $\epsilon$ is the UV cutoff.  After considering the finite part of eq. \eqref{EE-CFT-two disjoint interval-1}, and noticing that our background is conformally flat, one has
\footnote{Notice that the metric \eqref{metric-black brane-Kruskal} is conformally flat and the distances $d(a_{\pm} , b_{\pm} )$, i.e. eq. \eqref{d}, contain the warp factor $W(r)$ introduced in eq. \eqref{metric-black brane-Kruskal}.}
\bea
S_{\rm matter}^{\rm f} (\mathcal{R} \cup \mathcal{I}) = \frac{c}{3} \log \left( \frac{d(a_+,a_-) d(b_+,b_-) d(a_+,b_+) d(a_-,b_-)}{d(a_+,b_-) d(a_-,b_+)} \right).
\label{EE-CFT-two disjoint interval}
\eea 
Next, by applying eq. \eqref{d}, one obtains
\bea
S_{\rm matter}^{\rm f} (\mathcal{R} \cup \mathcal{I})  &=& \frac{c}{3} \Bigg[
\log \left( \frac{\beta^2 a^{z- \theta_e} \sqrt{f(a)}}{ \pi^2} \cosh \left( \frac{2 \pi t_a}{\beta} \right) \cosh \left( \frac{2 \pi t_b}{\beta} \right)\right)
\cr && \cr
&& \;\;\; + \log \left( \frac{ \cosh \left( \frac{2 \pi}{\beta} (b - r^*(a)) \right) - \cosh \left(  \frac{2 \pi}{\beta} (t_a - t_b) \right)}{ \cosh \left( \frac{2 \pi}{\beta} (b - r^*(a)) \right) + \cosh \left(  \frac{2 \pi}{\beta} (t_a + t_b) \right) }\right)
\Bigg].
\label{S-matter-CFT-islands}
\eea 
After adding the gravitational part, the generalized entropy becomes
\bea 
S_{\rm gen}  &=& \frac{V_d a^{d_e}}{2 G_{N,r}} + S_{\rm matter}^{\rm f}  (\mathcal{R} \cup \mathcal{I})
\cr&&\cr
&=& \frac{V_d a^{d_e}}{2 G_{N,r}} + 
 \frac{c}{3} \Bigg[
\log \left( \frac{\beta^2 a^{z- \theta_e} \sqrt{f(a)}}{ \pi^2} \cosh \left( \frac{2 \pi t_a}{\beta} \right) \cosh \left( \frac{2 \pi t_b}{\beta} \right)\right)
\cr && \cr
&& + \log \left( \frac{ \cosh \left( \frac{2 \pi}{\beta} (b - r^*(a)) \right) - \cosh \left(  \frac{2 \pi}{\beta} (t_a - t_b) \right)}{ \cosh \left( \frac{2 \pi}{\beta} (b - r^*(a)) \right) + \cosh \left(  \frac{2 \pi}{\beta} (t_a + t_b) \right) }\right)
\Bigg]
\cr && \cr
&=&
\frac{V_d a^{d_e}}{2 G_{N,r}} + 
\frac{c}{3} \Bigg[
\log \left( \frac{\beta^2 a^{z- \theta_e} \sqrt{f(a)}}{ \pi^2} \cosh \left( \frac{2 \pi t_a}{\beta} \right) \cosh \left( \frac{2 \pi t_b}{\beta} \right)\right) 
\cr && \cr && 
+ \log \left( \frac{1+ e^{-\frac{4 \pi }{ \beta} (b - r^*(a)) } - 2 e^{-\frac{2 \pi }{ \beta} (b - r^*(a)) } \cosh \left( \frac{2 \pi}{\beta} (t_a - t_b) \right) }{ 1+ e^{-\frac{4 \pi }{ \beta} (b - r^*(a)) } + 2 e^{-\frac{2 \pi }{ \beta} (b - r^*(a)) } \cosh \left( \frac{2 \pi}{\beta} (t_a + t_b) \right) }\right) \Bigg].
\label{S-gen-CFT-islands}
\eea 
Now we first find the entropy of Hawking radiation at early times, i.e. $t_{a,b} T \ll 1$. We assume that $t_a $ and $t_b$ are of the same order. Moreover, we consider the case where the endpoints of the island are close to the horizon, i.e. $a \approx r_h$ \cite{Hashimoto:2020cas}. 
\footnote{Notice that when $a \rightarrow r_h$, one has $r^*(a) \rightarrow - \infty$, and hence one may apply $\cosh \left( \frac{2 \pi}{\beta} (b - r^*(a)) \right) \approx \frac{1}{2} e^{ \frac{2 \pi}{\beta} (b - r^*(a)) }$.}
Therefore, one may apply the following approximations (See also \cite{Alishahiha:2020qza})
\bea
e^{\frac{2 \pi}{\beta} (b - r^*(a))}  &\gg &   e^{-\frac{2 \pi}{\beta} ( b- r^*(a)) } -2 \cosh \left( \frac{2 \pi}{\beta} (t_a - t_b) \right),
\cr && \cr 
e^{\frac{2 \pi}{\beta} (b - r^*(a))} &\gg &   e^{\frac{- 2 \pi}{\beta} ( b - r^*(a) } +2 \cosh \left( \frac{2 \pi}{\beta} (t_a + t_b) \right).
\label{approx-early times-2}
\eea 
In this case, by applying eq. \eqref{approx-early times-2}, one can expand the logarithmic term in \eqref{S-gen-CFT-islands} to second order and rewrite the equation as follows
\footnote{We keep terms of order $e^{-\frac{2 \pi}{\beta} (b - r^*(a))}$ and $e^{-\frac{4 \pi}{\beta} (b - r^*(a))}$ and omit higher order terms.}
\bea
S_{\rm gen}  &=& 
\frac{V_d a^{d_e}}{2 G_{N,r}} + 
\frac{c}{3} \Bigg[
\log \left( \frac{\beta^2 a^{z- \theta_e} \sqrt{f(a)}}{ \pi^2} \cosh \left( \frac{2 \pi t_a}{\beta} \right) \cosh \left( \frac{2 \pi t_b}{\beta} \right)\right) 
\cr && \cr
&& 
- 4 e^{-\frac{2 \pi }{ \beta} (b - r^*(a)) } \cosh \left( \frac{2 \pi t_a}{\beta} \right) \cosh \left( \frac{2 \pi t_b}{\beta} \right)
\cr && \cr
&& 
+ 2 e^{-\frac{4 \pi }{ \beta} (b - r^*(a)) } \left[ \cosh^2 \left( \frac{2 \pi (t_a + t_b)}{\beta} \right) - \cosh^2 \left( \frac{2 \pi (t_a- t_b)}{\beta} \right) \right] 
\Bigg].
\label{S-gen-CFT-islands-early-times}
\eea 
Next, by extremizing the above expression with respect to $a$, it is straightforward to verify that there is no real solution for $a$. Therefore, one may conclude that at early times, there are no islands, and hence $S_{\rm R}$ is given by eq. \eqref{S-R-CFT-no islands-2}.
\\Now we consider the entropy of Hawking radiation at late times, i.e. $t_{a,b} T \gg 1$. We again assume that $t_a$ and $t_b$ are of the same order and $a \approx r_h$. In this case, we can apply the following approximations (See also \cite{Hashimoto:2020cas,Alishahiha:2020qza})
\bea
e^{\frac{2 \pi}{\beta} (b - r^*(a))}  &\gg &   e^{-\frac{2 \pi}{\beta} ( b- r^*(a)) } -2 \cosh \left( \frac{2 \pi}{\beta} (t_a - t_b) \right),
\label{approx-late-times-2-1}
\\
2 \cosh \left( \frac{2 \pi}{\beta} (t_a + t_b) \right)  &\gg &  e^{\frac{2 \pi}{\beta} (b - r^*(a))}+ e^{\frac{- 2 \pi}{\beta} ( b - r^*(a) }.
\label{approx-late-times-2-2}
\eea 
Therefore, by applying eqs. \eqref{approx-late-times-2-1} and \eqref{approx-late-times-2-2}, one has
\footnote{It should be pointed out that the term $e^{-\frac{4 \pi }{ \beta} (b - r^*(a))}$ in eq. \eqref{S-gen-CFT-islands-late-times} gives a correction of order $G_{N,r}^2$ to the entropy of Hawking radiation $S_{\rm R}$. In the following, we calculate $S_{\rm R}$ to order $G_{N,r}^0$ (See e.g. eq. \eqref{S-R-CFT-island-late-times}), and hence one may omit this term similar to refs \cite{Hashimoto:2020cas,Alishahiha:2020qza}.}
\bea
S_{\rm gen} &=& 
\frac{V_d a^{d_e}}{2 G_{N,r}} +  \frac{c}{3} \Bigg[
\log \left( \frac{\beta^2 a^{z- \theta_e} \sqrt{f(a)}}{ \pi^2} \cosh \left( \frac{2 \pi t_a}{\beta} \right) \cosh \left( \frac{2 \pi t_b}{\beta} \right)\right) + \frac{2 \pi }{ \beta} (b - r^*(a)) 
\cr && \cr &&
- \log \left( 2 \cosh \left( \frac{2 \pi}{\beta } (t_a+t_b) \right)\right)
-2 e^{-\frac{2 \pi }{ \beta} (b - r^*(a)) } \cosh \left( \frac{2 \pi}{\beta} (t_a - t_b) \right)+ e^{-\frac{4 \pi }{ \beta} (b - r^*(a))}
\Bigg] 
\cr && \cr
 &=& 
\frac{V_d a^{d_e}}{2 G_{N,r}} +  \frac{c}{3} \Bigg[
\log \left( \frac{\beta^2 a^{z- \theta_e} \sqrt{f(a)}}{ 4 \pi^2} \right)
+ \frac{2 \pi }{ \beta} (b - r^*(a)) 
\cr && \cr &&\;\;\;\;\;\;\;\;\;\;\;\;\;\;\;\;\;\;\;
-2 e^{-\frac{2 \pi }{ \beta} (b - r^*(a)) } \cosh \left( \frac{2 \pi}{\beta} (t_a - t_b) \right)+ e^{-\frac{4 \pi }{ \beta} (b - r^*(a)) } 
\label{S-gen-CFT-islands-late-times}
\Bigg].
\eea 
Next, by extremizing $S_{\rm gen}$ with respect to $t_a$, one has
\bea
t_a = t_b.
\label{ta=tb-CFT}
\eea 
Then we plug eq. \eqref{ta=tb-CFT} into eq. \eqref{S-gen-CFT-islands-late-times} and extremize $S_{\rm gen}$ with respect to $a$. By expanding $\frac{\partial S_{\rm gen}}{\partial a}$ in powers of $\alpha = \sqrt{ \frac{a- r_h}{r_h}}$, one has
\bea
\frac{ \partial S_{\rm gen}}{ \partial a} &=&  \frac{ V_d r_h^{d_e-1} d_e }{2 G_{N,r}} + \frac{c \left( 2(z - \theta_e) - d_e  + 2 (d_e+z) e^{ \gamma- b (d_e+z) r_h^z + \psi \left( \frac{z}{d_e+z} \right)} \right)}{6 r_h} 
\cr && \cr 
&& - \frac{c}{3 r_h \alpha} \sqrt{d_e+z} e^{\frac{1}{2} \left( \gamma- b (d_e+z) r_h^z + \psi \left( \frac{z}{d_e + z} \right)\right)} + \mathcal{O} \left( \alpha \right) 
= 0.
\eea 
From which one can easily find the value of $a$ as follows
\bea
a =r_h + \left(\frac{ 2 c G_{N,r} }{3 V_d d_e} \right)^2 \frac{(d_e+z)}{r_h^{2d_e -1}}e^{\gamma - b (d_e+z) r_h^z + \psi \left( \frac{z}{ d_e +z} \right)} + \mathcal{O} \left( G_{N,r}^3 \right),
\label{a-CFT}
\eea 
where $\gamma=0.577$ is Euler's constant and $\psi ({\alpha})$ is the digamma function. Next, by plugging eqs. \eqref{ta=tb-CFT} and \eqref{a-CFT} into \eqref{S-gen-CFT-islands-late-times}, one can find the entropy of Hawking radiation as follows
\bea
S_{\mathcal{R}} &=& 2 S_{\rm th} + \frac{c}{6} \Bigg[
b (d_e+ z) r_h^{z} - \gamma - \psi \left( \frac{z}{d_e+ z} \right) 
\cr && \cr 
&& \;\;\;\;\;\;\;\;\;\;\;\;\;\;\;\;\;\;\;\;\;\;\;\;\; +  \log \left( \frac{16}{ (d_e+ z)^4 r_h^{2(z + \theta_e) }} \right) + \mathcal{O} \left( G_{N,r} \right)
\Bigg],
\label{S-R-CFT-island-late-times}
\eea 
where the first therm is twice the thermodynamic entropy $S_{\rm th}$ of a black brane. Notice that $S_{\rm R}$ is independent of time. Therefore, at late times the presence of the island leads $S_{\rm R}$ to saturate. In figure \ref{fig: S-R-CFT}, $S_R$ is plotted for different values of the exponents $z$ and $\theta$ of the black brane geometry. At late times, before the Page time, $S_{\rm R}$ has a linear growth (See eq. \eqref{S-R-CFT-no islands-late-times}). However, after the Page time, it is independent of time (See eq. \eqref{S-R-CFT-island-late-times}).
\\On the other hand, by equating eqs. \eqref{S-R-CFT-no islands-late-times} and \eqref{S-R-CFT-island-late-times}, the Page time is obtained as follows
\bea
t_{\rm page} &=&  \frac{ 3 V_d r_h^{d_e - z}}{c G_{N,r} (d_e+ z)} + \mathcal{O} \left( G_{N,r}^0 \right),
\cr&&\cr
&=& \frac{3}{\pi} \frac{S_{th}}{c T}.
\label{t-page-CFT}
\eea 
Therefore, the Page time is proportional to $\frac{S_{th}}{c T}$ and depends on both the exponents $z$ and $\theta$ of the black brane geometry. It should be pointed out that for $\theta =0$ and $z=1$ the black brane geometry becomes a d+2 dimensional planar AdS-Schwarzschild black hole. In this case, our results for $d=2$ reduce to those reported in section 5 of ref. \cite{Alishahiha:2020qza}. 
\footnote{Note that in ref. \cite{Alishahiha:2020qza}, one should turn off all of the higher derivative couplings in the action of the critical gravity to get an AdS-Schwarzschild black hole in Einstein gravity.}
Moreover, in the left panel of figure \ref{fig: t-Page-CFT}, $t_{\rm Page}$ is plotted as a function of $z$ for different values of $\theta$. It is observed that $t_{\rm Page}$ is a decreasing function of $z$.
Furthermore, in the right panel of figure \ref{fig: t-Page-CFT}, $t_{\rm Page}$ is plotted as a function of $\theta$ for some values of $z$. It is observed that $t_{\rm Page}$ is an increasing function of $\theta$. Form these diagrams, it is evident that for $\theta \leq 0$, the Page time is always smaller than that for the case $\theta=0$ and $z=1$. In other words, the entropy of Hawking radiation for a HV black brane with $\theta \leq 0$, saturates sooner than that for a planar AdS-Schwarzschild black hole. However, for positive values of $\theta$, by decreasing $z$, the Page time can become larger than that for the case $\theta=0$ and $z=1$.

\section{Entropy of Hawking Radiation: Matter HV QFT}
\label{Sec: Entropy of Hawking Radiation with Matter HV QFT}

In this section, we study the entropy of Hawking radiation for the case where the matter fields are described by a d+2 dimensional HV QFT which has a dual gravity that is a d+3 dimensional HV geometry at zero temperature. The metric of this geometry is simply obtained by setting $r_h=0$ in eq. \eqref{metric-BB} as follows \cite{Charmousis:2010zz}
\footnote{Note that to have a d+3 dimensional dual gravity we also replace $d$ with $d+1$ in eq.  \eqref{metric-HV-zero-temp}.}
\footnote{Some measures of quantum entanglement such as holographic entanglement entropy, mutual information and entanglement wedge cross section were studied for this background in the literature. See, e. g. \cite{Alishahiha:2012cm,Alishahiha:2015goa,MohammadiMozaffar:2015wnx,BabaeiVelni:2019pkw,Khoeini-Moghaddam:2020ymm,Dong:2012se,Tanhayi:2017wcd,Fischler:2012uv,Bueno:2014oua,Cavini:2019wyb}, which is not an exhaustive list.}
\bea
ds^2 =r^{- \frac{2 \theta_m}{d+1} } \left( - r^{2z_m} dt^2  + \frac{dr^2}{r^2} + r^2 \sum_{i=1}^{d+1} dx_i^2 \right).
\label{metric-HV-zero-temp}
\eea
We denote the dynamical and hyperscaling violation exponents of the corresponding HV QFT by $z_m$ and $\theta_m$, respectively to draw a distinction between them and the exponents $z$ and $\theta$ of the HV black brane. Moreover, for $z_m =1$ and $\theta_m=0$, the metric \eqref{metric-HV-zero-temp} becomes an $AdS_{d+3}$ spacetime which is dual to a $CFT_{d+2}$. Therefore, in this case, all of the results of this section should reduce to those in the previous section. Another important point is that since the HV QFT has a dual gravity, we can apply the RT formula \cite{Ryu:2006bv} for the calculation of the EE of matter fields as well as the entropy of Hawking radiation.

\subsection{Holographic Entanglement Entropy in a d+2 Dimensional HV QFT}
\label{Sec: Holographic Entanglement Entropy in a d+2 Dimensional HV QFT}

Now, we review the holographic entanglement entropy (HEE) for 
the metric \eqref{metric-HV-zero-temp} which is derived in ref. \cite{Dong:2012se}. We consider an entangling region in the shape of a strip on a constant time slice of the dual QFT whose width in the $x_1$ direction is $\ell$, and denote its lengths along the $x_{i  \neq 1}$ directions by $L$. It was observed that for $d = \theta_m $, the dual QFT has a Fermi surface and the HEE is given by \cite{Dong:2012se}
\bea
S = \frac{L^{d}}{2 \tilde{G}_N} \log \left( \frac{\ell}{\epsilon}\right),
\label{EE-strip-de=1}
\eea 
where $\tilde{G}_N$ is Newton's constant in d+3 dimensions, and $\epsilon$ is the cutoff in the dual QFT. 
\footnote{One might regard the HV QFT as an IR theory which can be obtained by a deformation of a CFT in the UV. In this sense, the cutoff $\epsilon$ should be considered as an effective cutoff 
$\epsilon_{\rm eff}$ 
\cite{Shaghoulian:2011aa}. 
When the HV QFT has a Fermi surface, one has $\epsilon_{\rm eff} \propto k_F^{-1}$, where $k_F$ is the radius of the corresponding Fermi surface \cite{Shaghoulian:2011aa}. In other words, the gravity with the metric in eq. \eqref{metric-HV-zero-temp} is well defined only above a dynamical scale $r_F$ i.e. $r > { r_F}$ \cite{Shaghoulian:2011aa,Dong:2012se} (Notice that our radial coordinate $r$ is related to that in ref. \cite{Dong:2012se} by $r_{\rm here} = \frac{1}{r_{\rm there}}$ ).
In this manner, the cutoff dependent terms in the HEE are IR effects and not divergent in the UV. We would like to thank Edgar Shaghoulian very much for his helpful comments on this topic.}
On the other hand, for $d \neq \theta_m$, the HEE is given by \cite{Dong:2012se}
\bea
S =  \frac{L^{d}}{2 \tilde{G}_N (d - \theta_m)} \left[ \frac{1}{\epsilon^{d -\theta_m} } - \Upsilon^{d - \theta_m +1} \left(\frac{2}{\ell}\ \right) ^{d - \theta_m} \right],
\label{EE-strip-de neq 1}
\eea 
where 
\bea
\Upsilon = \frac{\sqrt{\pi} \Gamma \left(\frac{d - \theta_m +2}{2 (d - \theta_m +1)} \right)}{\Gamma \left(\frac{1}{2 ( d - \theta_m) +1 }\right)}.
\label{Upsilon}
\eea 
In the following, we renormalize Newton's constant to absorb the cutoff dependent terms in the HEE.
For $d = \theta_m$, from eq. \eqref{EE-strip-de=1}, one can renormalize Newton's constant as follows 
\bea
\frac{1}{4 G_{N,r}} = \frac{1}{4 G_N} + \log \epsilon.
\label{G_N-r-d = thetam}
\eea
However, for $d \neq \theta_m$, from eq. \eqref{EE-strip-de neq 1}, one has
\footnote{Notice that we set the dynamical scale $r_F$ to one. Otherwise, it is written as $ \frac{1}{4 G_{N,r}} = \frac{1}{4 G_N} - \frac{1}{r_F^{\theta_m} \epsilon^{d- \theta_m}}$.}
\bea
\frac{1}{4 G_{N,r}} = \frac{1}{4 G_N} - \frac{1}{\epsilon^{d- \theta_m}}.
\label{G_N-r-d neq thetam}
\eea
Notice that for $\theta_m=0$, it reduces to eq. \eqref{G_N-r-CFT}. Therefore, we only consider those parts
of $S_{\rm matter} ( \mathcal{R} \cup \mathcal{I})$ which are independent of the cutoff $\epsilon$ in what follows and again rewrite $S_{\rm gen}$ as eq. \eqref{S-gen-ren}. Moreover, as mentioned in section \ref{Sec: Entropy of Hawking Radiation with Matter CFT}, due to the translational symmetry of the metric \eqref{metric-BB} along the traverse directions $x^i$, one might effectively work in two dimensions parametrized by the coordinates $t$ and $r$. Therefore, we need to know the HEE of an interval in a two dimensional HV QFT to calculate the entropy of Hawking radiation.
\footnote{Note that to have a two-dimensional HV QFT, one should set $d=0$ in eq. \eqref{metric-HV-zero-temp}.}
In this case, the cutoff independent parts of the HEE are as follows
\begin{itemize}
	\item For $\theta_m =0$, by applying eq. \eqref{EE-strip-de=1}, one has
	\bea
	S^{\rm f} = \frac{A}{3}  \log \ell,
	\label{EE-de=1-2d}
	\eea 
	where 
	\footnote{Recall that we set the AdS radius $R$ to one.}
	\bea 
	A= \frac{3}{2 \tilde{G}_N}.
	\label{A}
	\eea 
	Notice that, for $\theta_m=0$ and $z_m=1$, the HV $QFT_2$ becomes a holographic $CFT_2$, and hence "A" is equal to the corresponding central charge $c$ \cite{Brown:1986}.
	\item For $\theta_m  \neq 0$, by applying eq. \eqref{EE-strip-de neq 1}, one has
	\bea
	S^{\rm f} = \frac{ A}{ 3 \theta_m } \Upsilon_0^{1 - \theta_m } \left(\frac{ \ell}{2}\ \right) ^{ \theta_m},
	\label{EE-de neq 1-2d}
	\eea 
\end{itemize}
where
\bea
\Upsilon_0 = \frac{\sqrt{\pi} \Gamma \left(\frac{2- \theta_m}{2 (1- \theta_m)} \right)}{\Gamma \left(\frac{1}{2 ( 1- \theta_m)}\right)}.
\label{Upsilon-m}
\eea 
Notice that in this case $S^{\rm f} $ is negative (positive) when $\theta_m$ is negative (positive). Moreover, we impose the constraint $d +1 > \theta_m $ which for $d=0$ leads to 
\bea
\theta_m <1.
\label{theta-less-1}
\eea 
In the next section, we calculate the entropy of Hawking radiation for the case where the matter QFT is a d+2 dimensional HV QFT. We study the two cases $\theta_m=0$ and $\theta_m \neq 0$, separately. Moreover, to have a positive entropy of Hawking radiation when there are no islands and $\theta_m \neq 0$, the finite part of the EE of matter has to be positive, i.e. $\theta_m >0$. Combining this inequality with eq. \eqref{theta-less-1} leads to the constraint $0 < \theta_m < 1$. 

\subsection{Entropy of Hawking Radiation for $\theta_m =0$}
\label{Sec: theta = 0}

In this section, we study the case where the hyperscaling violation exponent $\theta_m$ of the matter fields is zero. We first assume that there are no islands and calculate the entropy of Hawking radiation. We will see that it grows linearly in time and violates the unitarity, i.e. eq. \eqref{S-R leq-SBH-two sided}. Therefore, one needs to include the contribution of an island to stop this growth and saturate the entropy of Hawking radiation.

\subsubsection{No Islands}
\label{Sec: No Islands- theta-0}

In this case, there are only the radiation regions $ \mathcal{R} = \mathcal{R}_- \cup \mathcal{R}_+$ (See the left panel of figure \ref{fig: Penrose}). Similar to the previous section, we assume that the whole state on the Cauchy slice is pure. Therefore, one can calculate the EE of the matter fields on the interval $[b_- , b_+]$ instead of that on the region $\mathcal{R}$, and hence the entropy of Hawking radiation is simply given by
\bea
S_{ \rm R} = S_{\rm matter}^{\rm f} \left( \mathcal{R} \right) = S_{\rm matter}^{\rm f} (b_+ ,b_-).
\label{S-R-theta-0-no island-1}
\eea 
Next, by applying eqs. \eqref{d} and \eqref{EE-de=1-2d}, one can rewrite eq. \eqref{S-R-theta-0-no island-1} as follows
\bea
S_{ \rm R} = \frac{A}{3} \log d(b_+, b_-) 
= \frac{A}{3} \log  \left( \frac{\beta}{\pi} \cosh \left( \frac{2 \pi t_b}{\beta}\right) \right).
\label{S-R-theta-0-no islands-2}
\eea 
It grows quadratically in time at early times, i.e. $t_b T \ll 1$,
\bea
S_{ \rm R} = \frac{A}{3} \left[ \frac{(d_e+z)^2 r_h^{2z}}{8} t_b^2 + \log \left( \frac{4}{ ( d_e+z)  r_h^z }\right) \right].
\label{S-R-theta-0-no islands-early-times}
\eea 
On the other hand, it grows linearly in time at late times, i.e. $t_b T \gg 1$,
\bea
S_{ \rm R} \simeq \frac{A (d_e+ z) r_h^z}{6} t_b,
\label{S-R-theta-0-no islands-late-times}
\eea 
which again will exceed the coarse-grained entropy of the two black branes. In the next section, we will see that by adding the contribution of an island to the entropy of  Hawking radiation, it saturates after the Page time.

\subsubsection{With Islands}
\label{Sec: With Islands- theta-0}

\begin{figure}
	\begin{center}
		\includegraphics[scale=0.85]{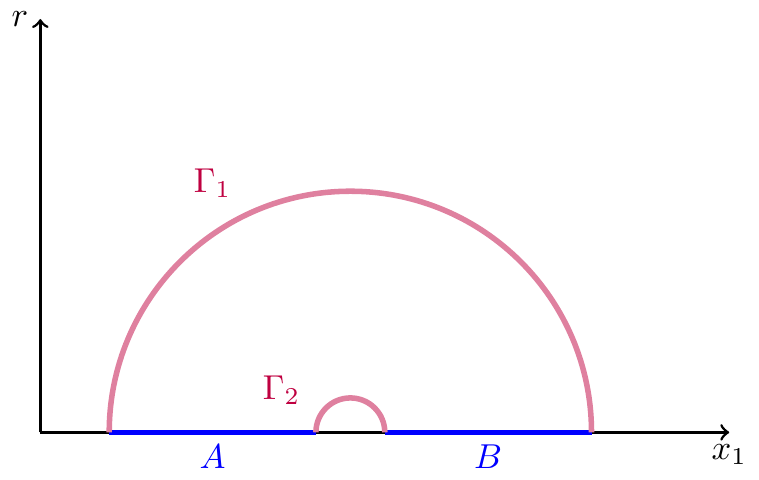}
		\includegraphics[scale=0.85]{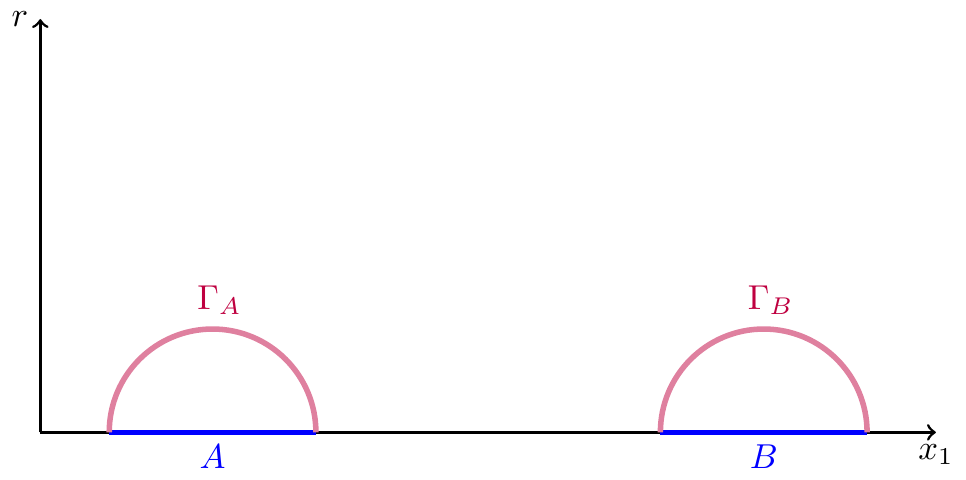}
	\end{center}
	\caption{An illustration of the Ryu-Takayanagi (RT) surfaces indicated in purple for the calculation of HEE when the entangling region is composed of two disjoint intervals $A$ and $B$ along the spacial direction $x_1$. Here $r$ is the radial coordinate in the dual gravity. {\it Left}) Connected configuration: each RT surface starts from an endpoint of one of the intervals and ends on an endpoint of another interval. {\it Right}) Disconnected configuration: each RT surface starts from one endpoint of an interval and ends on another endpoint of the same interval. It should be emphasized that the RT surfaces are not necessarily half circles and this diagram is schematic. 
	}
	\label{fig: RT-surfaces}
\end{figure}

In this section, 
we assume that there is an island $\mathcal{I}$ whose endpoints 
are located outside the event horizon (See the right panel of figure \ref{fig: Penrose}). By assuming that the state on the whole Cauchy slice is pure, one needs to calculate the EE of matter fields on the two disjoint intervals $[b_-,a_-] \cup [a_+ , b_+]$. Since the matter HV QFT has a holographic gravitational theory, one may apply the holographic prescription of ref. \cite{Headrick:2010zt} to calculate the EE of the two disjoint intervals. In this case, it is given by
\bea
S_{[b_-,a_-] \cup [a_+ , b_+]} = {\rm Min} \left( S_{\rm con} , S_{\rm dis} \right) ,
\label{S-two-interval-1}
\eea 
where 
$S_{\rm con}$
and 
$S_{\rm dis}$
are the HEE of the connected and disconnected configurations for the corresponding RT surfaces, respectively. Recall that in the connected configuration, each RT surface starts from an endpoint of one of the intervals and ends on an endpoint of another interval (See the left panel of figure \ref{fig: RT-surfaces}). However, in the disconnected configuration, each RT surface starts from an endpoint of one interval and ends on another endpoint of the same interval (See the right panel of figure \ref{fig: RT-surfaces}). From figure \ref{fig: RT-surfaces}, it is evident that for the intervals $A = [b_-,a_-]$ and $B= [a_+ , b_+]$, one can find
\bea
S_{\rm con}  &=& S(a_-,a_+) + S(b_-,b_+),
\cr && \cr
S_{\rm dis} &=& S(b_-,a_-) + S(a_+,b_+).
\label{EE-con-dis}
\eea 
In the following, we consider two regimes: early times and late times. 
At early times, i.e. $t_{a,b} T \ll 1$, the two endpoints $a_{\pm}$ of the island are close to each other. In this case, the lengths of the two intervals $[b_-,a_-]$ and $[a_+ , b_+]$ are much larger than the distance between them. Therefore, the EE is given by that of the connected RT surfaces (See the left panel of figure \ref{fig: RT-surfaces})
\bea
S_{\rm matter} (\mathcal{R} \cup \mathcal{I}) = S_{\rm con} = S_{\rm matter} (a_-,a_+) + S_{\rm matter} (b_-,b_+).
\label{S-matter-theta-0-islands-early-times-1}
\eea 
Next, by plugging eq. \eqref{d} and \eqref{EE-de=1-2d} into \eqref{S-matter-theta-0-islands-early-times-1}, one can find the finite part of the EE of the matter as follows
\bea
S_{\rm matter}^{\rm f} (\mathcal{R} \cup \mathcal{I})  &=&  \frac{A}{3} \log \left( d(a_-,a_+) d(b_-,b_+) \right)
\cr && \cr
&=& \frac{A}{3} \log \left( \frac{\beta^2 a^{z - \theta_e} \sqrt{f(a)}}{ \pi^2} \cosh \left( \frac{2\pi t_a}{\beta} \right) \cosh \left( \frac{2\pi t_b}{\beta}\right)\right).
\label{S-matter-theta-0-islands-early-times-2}
\eea 
Therefore the generalized entropy is given by
\bea
S_{\rm gen} &=& \frac{V_d a^{d_e}}{2 G_{N,r}} + S_{\rm matter}^{\rm f}  (\mathcal{R} \cup \mathcal{I})
\cr&&\cr
&=& \frac{V_d a^{d_e}}{2 G_{N,r}} + \frac{A}{3} \log \left( \frac{\beta^2 a^{z - \theta_e} \sqrt{f(a)}}{ \pi^2} \cosh \left( \frac{2\pi t_a}{\beta} \right) \cosh \left( \frac{2\pi t_b}{\beta}\right)\right).
\label{S-gen-theta-0-islands-early-times}
\eea 
Now, one can extremize $S_{\rm gen}$ with respect to $a$. If one assumes that $a \approx r_h$ and expands $\frac{\partial S_{\rm gen}}{\partial a}$ in powers of $\alpha = \sqrt{\frac{a-r_h}{r_h}}$, one obtains
\bea
\frac{\partial S_{\rm gen}}{\partial a} = \frac{V_d d_e r_h^{d_e-1} }{2 G_{N,r}} + \frac{A(3 z -1 - d_e - 4 \theta_e )}{12 r_h} + \frac{A}{6 r_h \alpha^2} + 
\mathcal{O} ( \alpha^2) =0.
\eea 
Then, it can be easily solved to find
\bea
a= r_h - \frac{A G_{N,r}}{3 d_e V_d r_h^{d_e-1}} + \mathcal{O} (G_{N,r}^2).
\label{a-theta-0-early-times}
\eea 
Next, by plugging eq. \eqref{a-theta-0-early-times} into \eqref{S-gen-theta-0-islands-early-times} and extremizing $S_{\rm gen}$ with respect to $t_a$, one simply obtains $t_a =0$. It is straightforward to check that for these values of $a$ and $t_a$, the generalized entropy is imaginary. Therefore, one may conclude that at early times there are no islands and the entropy of Hawking radiation is given by eq. \eqref{S-R-theta-0-no islands-2}.
\\On the other hand, at late times, i.e. $t_{a,b} T \gg 1$, the two endpoints of the island are very far from each other. Therefore, the lengths of the two intervals $[b_-,a_-]$ and $[a_+ , b_+]$ are much smaller than their distance. In this case, the EE is given by that of the disconnected RT surfaces (See the right panel of figure \ref{fig: RT-surfaces})
\footnote{It should be pointed out that this prescription were also applied in refs. \cite{Almheiri:2019yqk,Penington:2019kki,Almheiri:2019qdq,Hartman:2020swn} for two-dimensional eternal black holes at late times.}
\bea
S_{\rm matter} (\mathcal{R} \cup \mathcal{I}) = S_{\rm dis} = S_{\rm matter} (a_+,b_+) + S_{\rm matter} (a_-,b_-).
\label{S-matter-theta-0-islands-late-times}
\eea 
Next, by plugging eqs. \eqref{d} and \eqref{EE-de=1-2d} into \eqref{S-matter-theta-0-islands-late-times}, one has
\bea
S_{\rm matter}^{\rm f} (\mathcal{R} \cup \mathcal{I}) &=& \frac{ A}{3} \log \left( d(a_+,b_+) d(a_-,b_-) \right)
\\
&=& \frac{A}{3}
\log \Bigg[ \frac{\beta^2 a^{z - \theta_e} \sqrt{f(a)}}{ 2 \pi^2} \left( \cosh \left( \frac{2 \pi}{ \beta} (b - r^*(a)) \right) - \cosh \left( \frac{2 \pi}{ \beta} (t_a - t_b) \right)\right) \Bigg].
\nonumber
\eea 
At late times, by assuming that $a \approx r_h$ and $t_a \approx t_b$, one can apply the approximation given in eq. \eqref{approx-late-times-2-1}.
After adding the gravitational part, one has
\bea
S_{\rm gen} 
&=&  
\frac{V_d a^{d_e}}{2 G_{N,r}} + \frac{A}{3} \Bigg[
\log \left( \frac{ \beta^2 a^{z - \theta_e} \sqrt{f(a)}}{ 4 \pi^2}\right) + \frac{ 2 \pi}{ \beta} \left( b - r^*(a) \right)
\cr && \cr
&& \;\;\;\;\;\;\;\;\;\;\;\;\;\;\;\;\;\;\;\;\;\;  + e^{\frac{2 \pi}{\beta} ( r^*(a) -b)} \left( e^{\frac{2 \pi}{\beta} ( r^*(a) -b)} - 2 \cosh \left( \frac{2 \pi }{ \beta} (t_a - t_b) \right) \right)
\Bigg].
\label{S-gen-islands-theta=0}
\eea 
Next, by extremizing the generalized entropy with respect to $t_a$, 
one easily finds 
\bea
t_b=t_a.
\label{ta-tb-theta-0}
\eea
Then one can extremize $S_{\rm gen}$ with respect to $a$ and expand $\frac{\partial S_{\rm gen}}{\partial a}$ in powers of $\alpha = \sqrt{ \frac{a- r_h}{r_h}}$ as follows
\bea
\frac{ \partial S_{\rm gen}}{ \partial a} &=& \frac{ V_d r_h^{d_e-1} d_e }{2 G_{N,r}} + \frac{A \left( 2(z - \theta_e) - d_e  + 2 (d_e+z) e^{ \gamma- b (d_e+z) r_h^z + \psi \left( \frac{z}{d_e+z} \right)} \right)}{6 r_h} 
\cr && \cr 
&& 
\;\;\;\;\;\;\;\; - \frac{A}{3 r_h \alpha} \sqrt{d_e+z} e^{\frac{1}{2} \left( \gamma- b (d_e+z) r_h^z + \psi \left( \frac{z}{d_e + z} \right)\right)} + \mathcal{O} \left( \alpha \right) 
= 0.
\eea 
The above equation can be easily solved to find the value of $a$ as follows
\bea
a =r_h + \left(\frac{ 2 A G_{N,r} }{3 V_d d_e} \right)^2 \frac{(d_e+z)}{r_h^{2d_e -1}}e^{\gamma - b (d_e+z) r_h^z + \psi \left( \frac{z}{ d_e +z} \right)} + \mathcal{O} \left( G_{N,r}^3 \right).
\label{a-theta-0}
\eea 
At the end, by plugging eqs. \eqref{ta-tb-theta-0} and \eqref{a-theta-0} into \eqref{S-gen-islands-theta=0}, one can obtain the entropy of Hawking radiation as follows
\bea
S_{\mathcal{R}} & = & 2 S_{\rm th} + \frac{A}{6} \Bigg[
b (d_e+ z) r_h^{z} - \gamma - \psi \left( \frac{z}{d_e+ z} \right) 
\cr && \cr
&& \;\;\;\;\;\;\;\;\;\;\;\;\;\;\;\;\;\;\;\;\;\;\; +  \log \left( \frac{16}{ (d_e+ z)^4 r_h^{2(z + \theta_e) }} \right) + \mathcal{O} \left( G_{N,r} \right)
\Bigg].
\label{S-R-theta-0-island}
\eea 
The first term is twice the thermal entropy $S_{\rm th}$ of the black brane. Moreover, $S_{\rm R}$ is independent of time. Therefore, the presence of the island at late times leads $S_{\rm R}$ to become a constant. 
\\Furthermore, by equating eqs. \eqref{S-R-theta-0-no islands-late-times} and \eqref{S-R-theta-0-island}, one can obtain the Page time as follows
\bea
t_{\rm page} &=& \frac{ 3 V_d r_h^{d_e - z}}{A G_{N,r} (d_e+ z)} + \mathcal{O} \left( G_{N,r}^0 \right),
\cr&&\cr
&=& \frac{3}{\pi} \frac{S_{th}}{A T}.
\label{t-page-theta-0}
\eea 
Therefore, the Page time is proportional to $\frac{S_{th}}{A T}$ and depends on both the exponents $z$ and $\theta$ of the black brane geometry. Moreover, it is independent of the exponent $z_m$ of the matter fields. It should be pointed out that for $z_m=1$ and $\theta_m =0$, the HV $QFT_{d+2}$ reduces to a $CFT_{d+2}$. In this case, $A= \frac{3 R}{2 G_N}$ becomes equal to the central charge $c$ of the CFT. Therefore, it is expected that all of the results in this section reduces to those in section \ref{Sec: Entropy of Hawking Radiation with Matter CFT}, if one sets $z_m=1$. On the other hand, we observed that the EE of matter fields, and hence $S_{\rm R}$ are independent of the dynamical exponent $z_m$ of the matter fields. Therefore, all of the results in this section for $z_m \neq 1$, are the same as those for the matter CFT if one replaces $A$ with $c$. Therefore, the Page curve and Page time are again given by figures \ref{fig: S-R-CFT} and \ref{fig: t-Page-CFT}, respectively if one replaces $c$ with $A$.
\subsection{Entropy of Hawking Radiation for $\theta_m \neq 0$}
\label{Sec: Entropy of Hawking Radiation for theta-m neq 0}

In this section, we study the case where the hyperscaling violation exponent $\theta_m$ of matter fields is non-zero. We first study the case where there is no island. Next, we consider the effect of the presence of an island on the entropy of Hawking radiation. 

\subsubsection{No Islands}
\label{Sec: No Islands- theta-m-neq-0}

When there are no islands, by applying eq. \eqref{EE-de neq 1-2d}, one can write
\bea
S_{\mathcal{R}}= S_{\rm matter}^{f} (b_-,b_+) = \frac{A}{3 \theta_m} \Upsilon_0^{1 - \theta_m} \left( \frac{\beta}{ 2 \pi} \cosh \left( \frac{2 \pi t_b}{ \beta} \right) \right)^{\theta_m},
\label{S-R-theta neq 0-no island} 
\eea 
where $\theta_m$ is the hyperscaling violating exponent of the matter fields and $\Upsilon_0$ is defined by eq. \eqref{Upsilon-m}.
Notice that for negative values of $\theta_m$, $S_{\mathcal{R}}$ is negative. Therefore, in the following we restrict ourselves to the case $ 0 <\theta_m < 1$. At  early times, one has
\bea
S_{\mathcal{R}} = \frac{A}{3 \theta_m} \Upsilon_0^{1 - \theta_m} \left( \frac{2}{(d_e +z) r_h^z} \right)^{\theta_m} \left( 1 + \frac{1}{8} \theta_m (d_e +z)^2 r_h^{2 z} t_b^2 \right).
\label{S-R-theta neq 0-no island-early-times}
\eea 
Thus, $S_{\mathcal{R}}$ grows quadratically in time. On the other hand, at late times, one obtains
\bea
S_{\mathcal{R}} = \frac{A}{3 \theta_m} \Upsilon_0^{1 - \theta_m} \left( \frac{1}{ (d_e+z) r_h^z} \right)^{\theta_m} e^{\frac{(d_e+z) \theta_m r_h^z t_b}{2}}.
\label{S-R-theta neq 0-no island-late-times}
\eea 
Notice that $d_e + z >0$ and $0< \theta_m <1$, and hence the entropy of Hawking radiation grows exponentially in time. This behavior is in contrast to the usual linear growth which were previously observed for flat or AdS black holes in the literature, e.g. refs. \cite{Hashimoto:2020cas,Alishahiha:2020qza}.

\subsubsection{With Islands}
\label{Sec: With Islands- theta-neq-0}

\begin{figure}
	\begin{center}
		\includegraphics[scale=0.24]{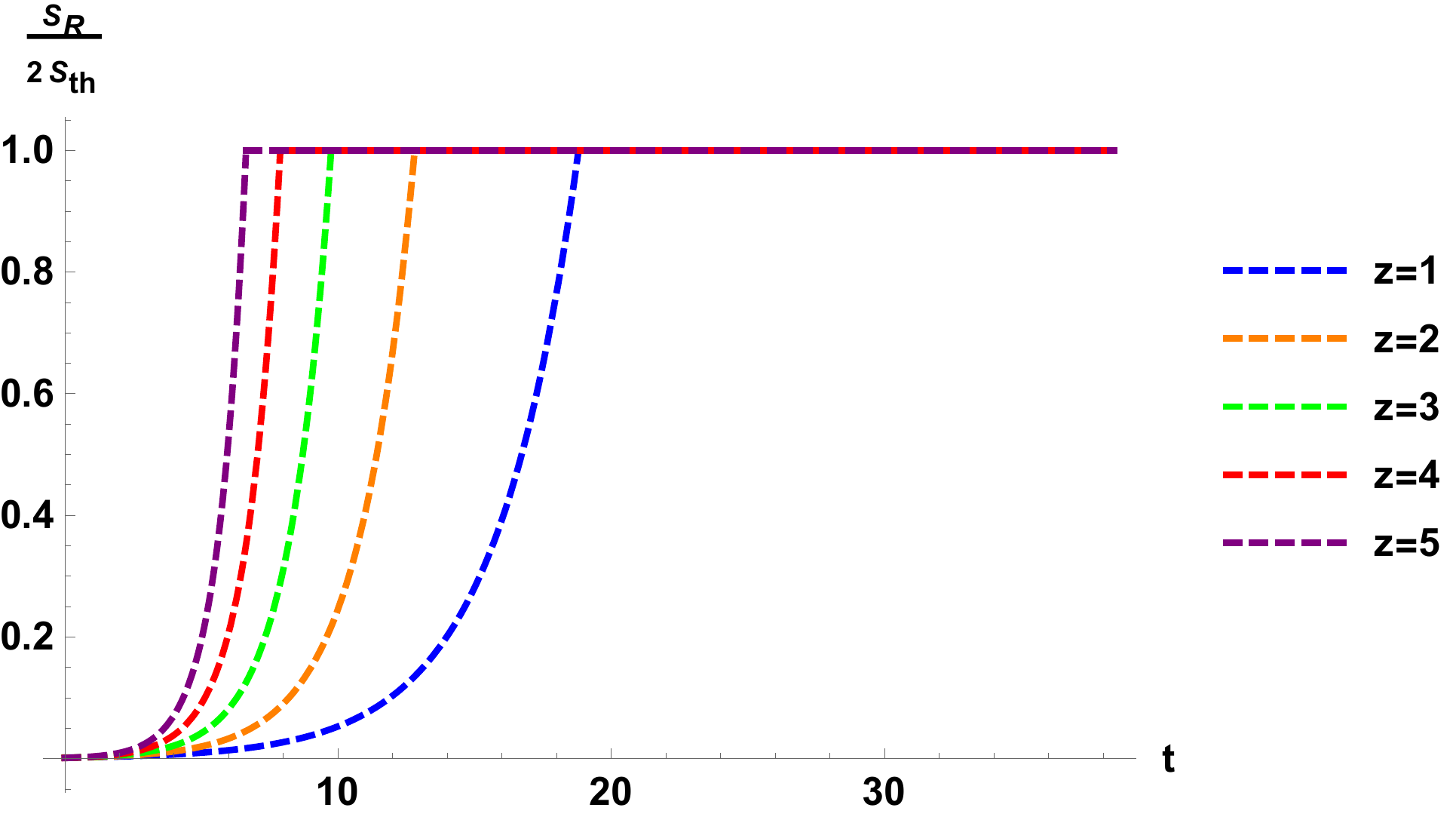}
		\includegraphics[scale=0.24]{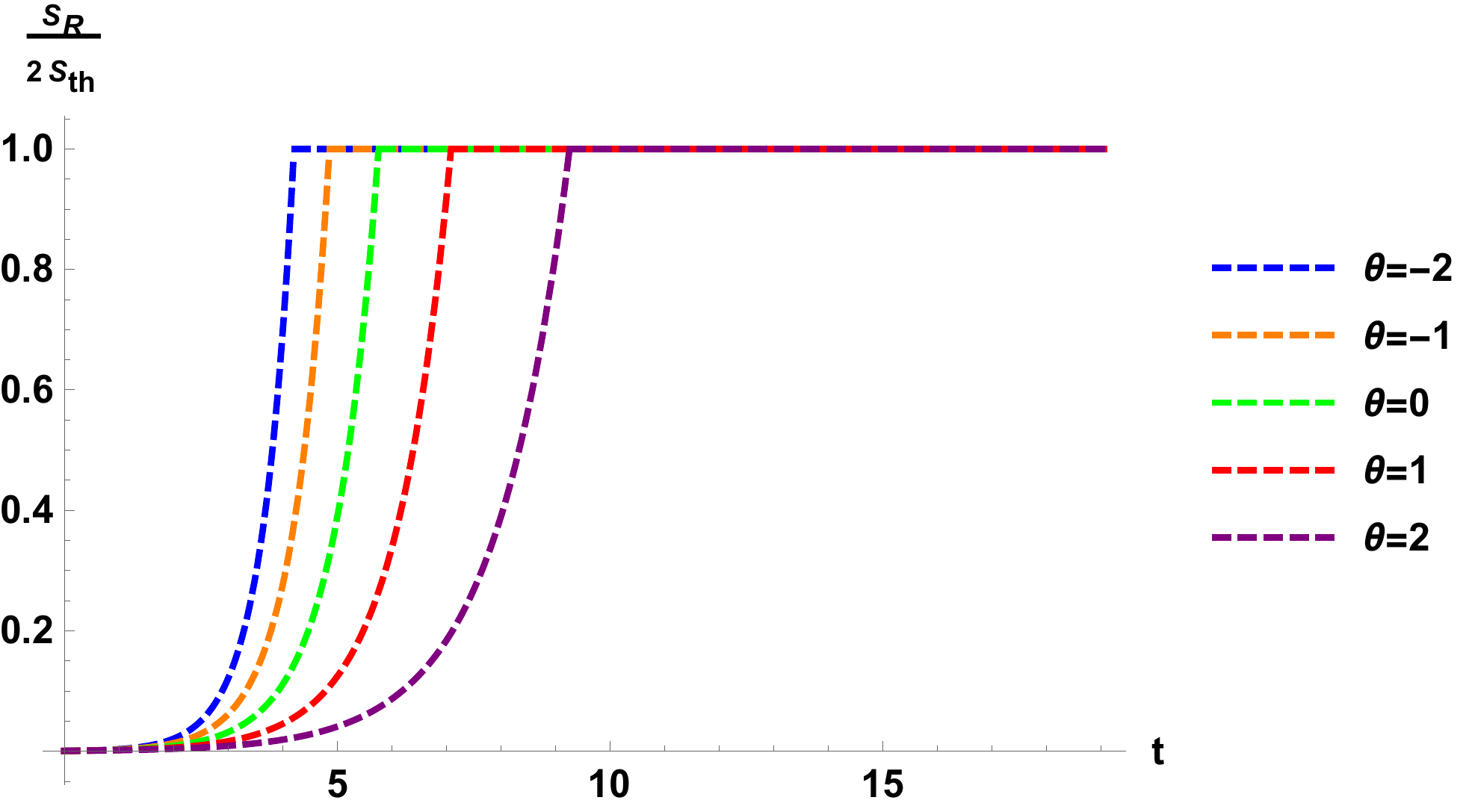}
		\includegraphics[scale=0.24]{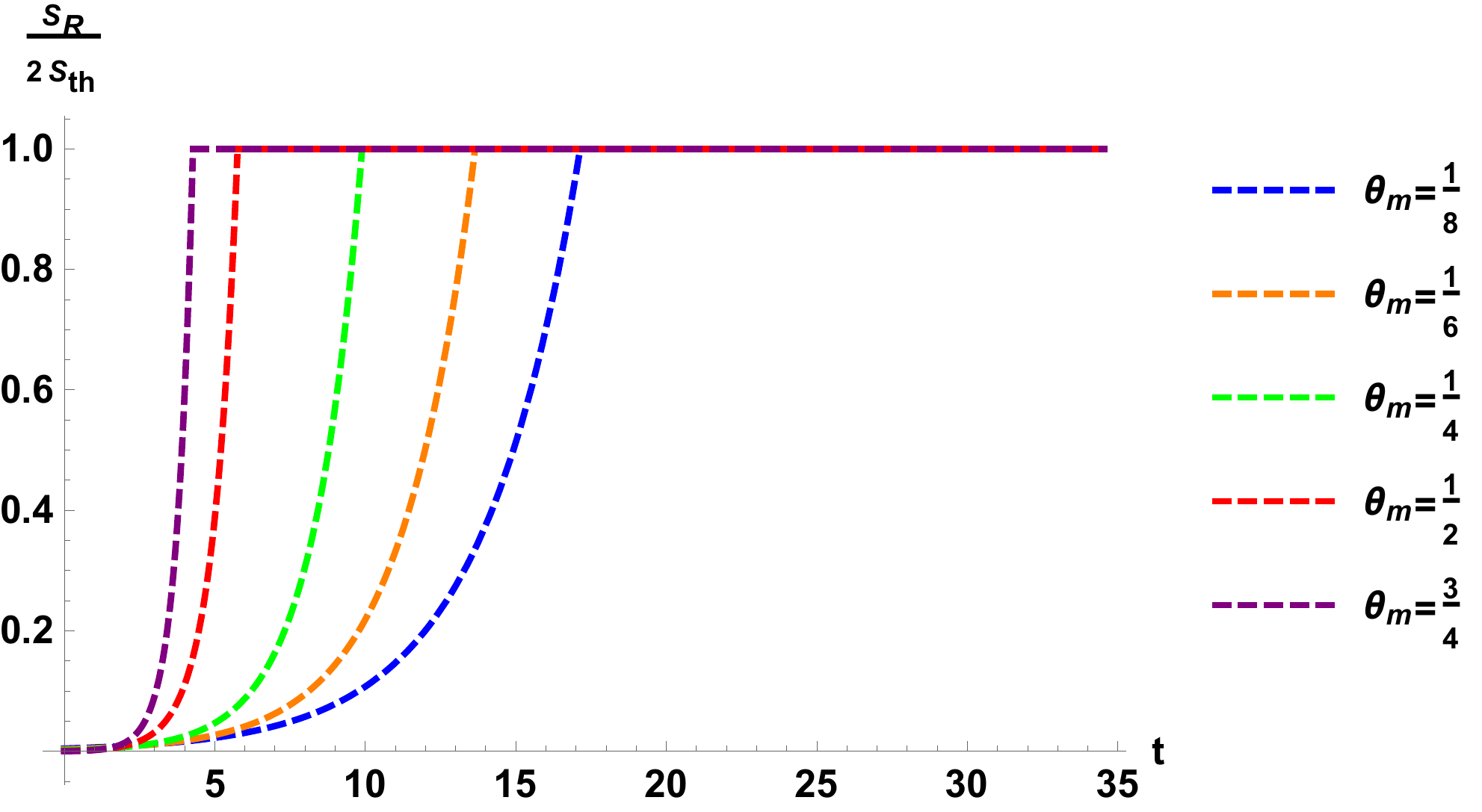}
	\end{center}
	\caption{
		Entropy of Hawking radiation $S_R$ when the matter is described by a HV $QFT_{d+2}$ with $\theta_m \neq 0$ as a function of $t$: {\it Left}) when $d=2$, $ \theta = 1$ and $\theta_m=\frac{1}{3}$ for some values of $z$. {\it Middle}) when $d=3$, $z=2$ and $\theta_m=\frac{1}{2}$ for different values of $\theta$. {\it Right}) when $d=3$, $z=1$ and $\theta=-1$ for some values of $\theta_m$. Here we set $A=V_d=r_h=1$ and $G_{N,r} =0.001$. It should be pointed out that we omit the correction of order $\mathcal{O}(G_{N,r}^0)$ in eq. \eqref{S-R-theta neq 0-islands-late-times}.
	}
	\label{fig: S-R-theta neq 0}
\end{figure}

\begin{figure}
	\begin{center}
		\includegraphics[scale=0.37]{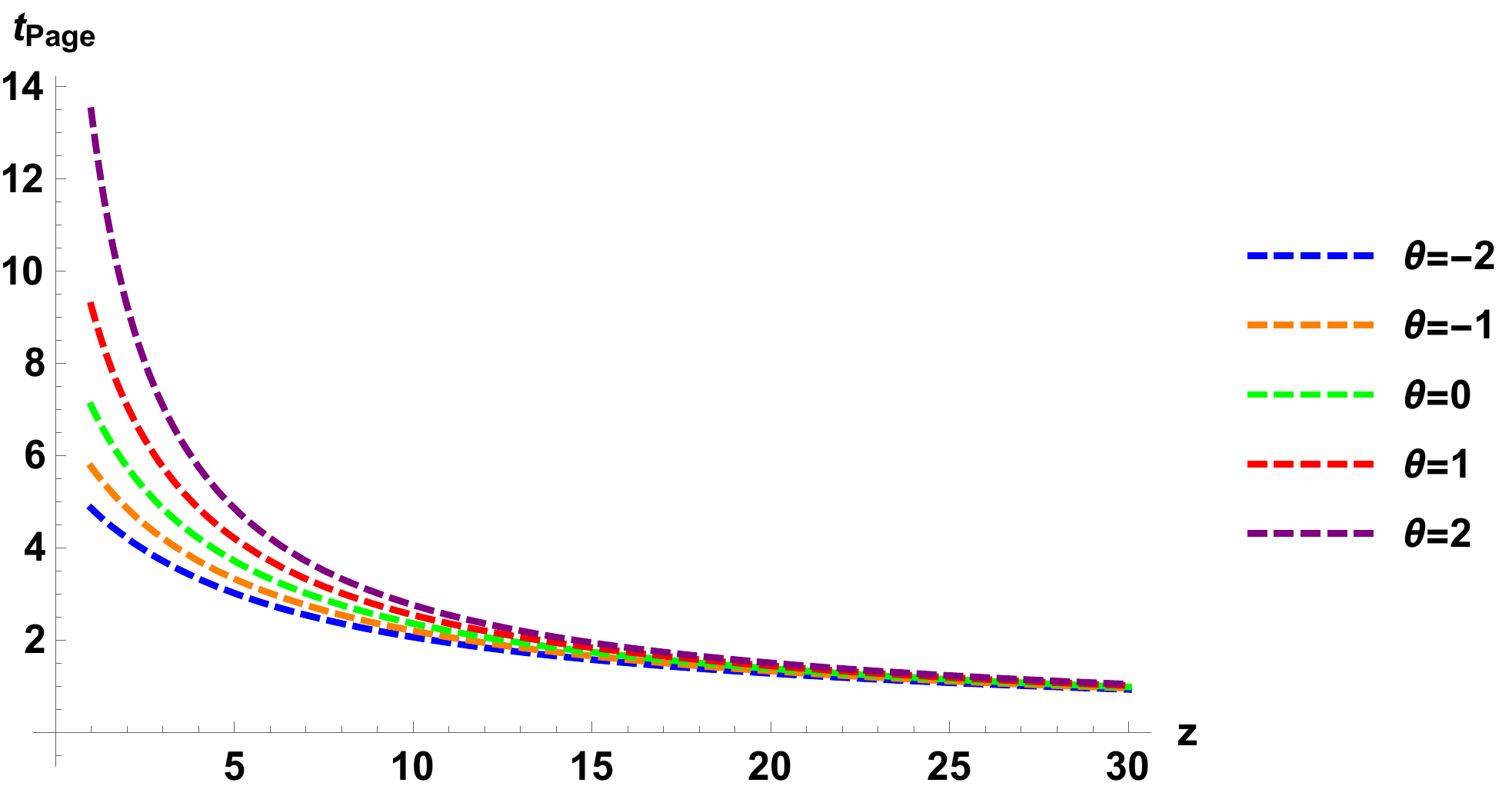}
				\hspace{0.2cm}
		\includegraphics[scale=0.37]{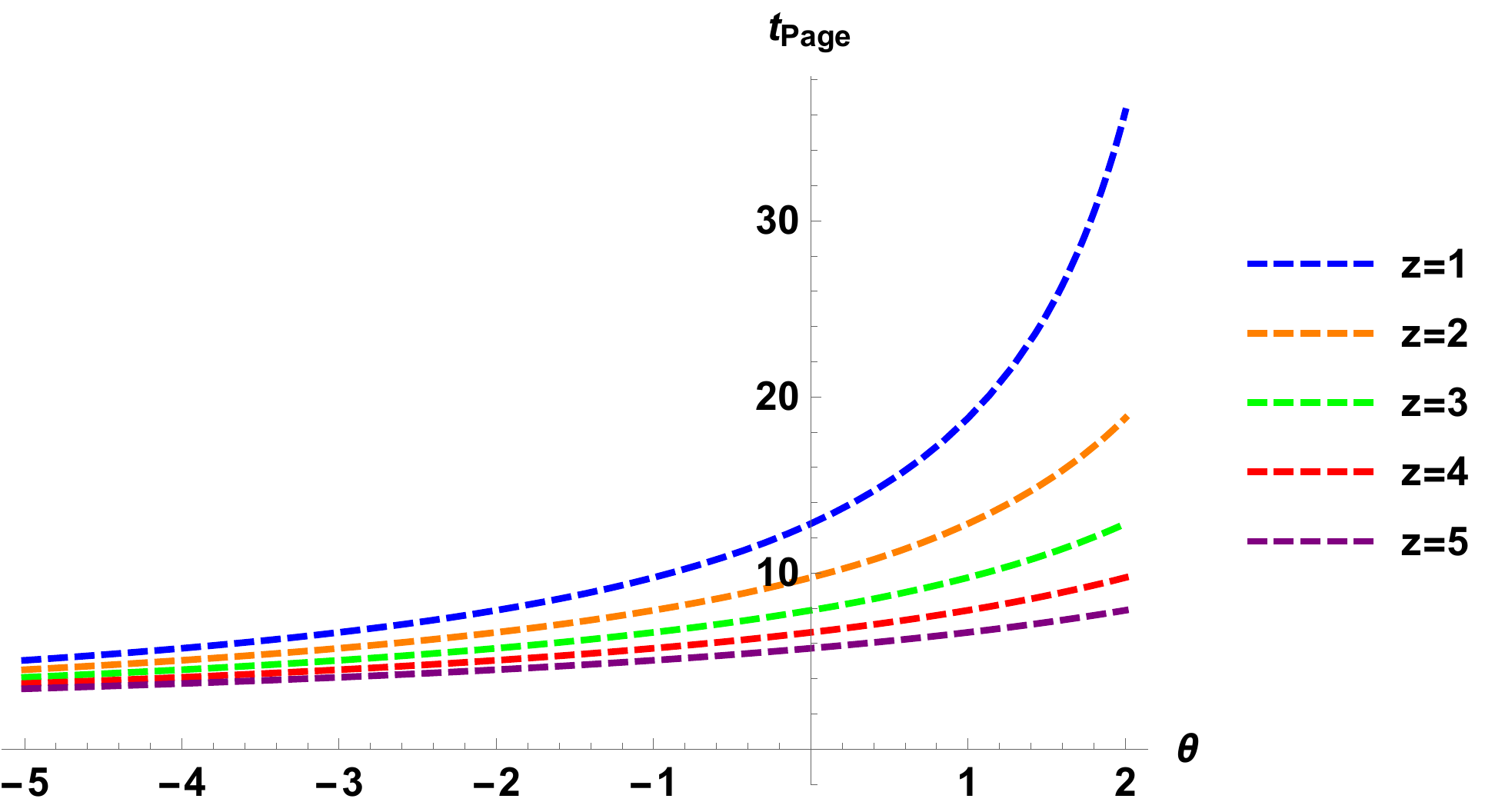}
	\end{center}
	\caption{
		Page time when the matter is described by a HV $QFT_{d+2}$ with $\theta_m \neq 0$ as a function of: {\it Left}) $z$ for $d=3$, $\theta_m=\frac{1}{2}$ and different values of $\theta$. {\it Right}) $\theta$ for $d=2$, $\theta_m=\frac{1}{3}$ and different values of $z$. Here we set $A=V_d=r_h=1$ and $G_{N,r} =0.001$.
		}
	\label{fig: t-page-thetam neq 0-z-theta}
\end{figure}

\begin{figure}
	\begin{center}
		\includegraphics[scale=0.37]{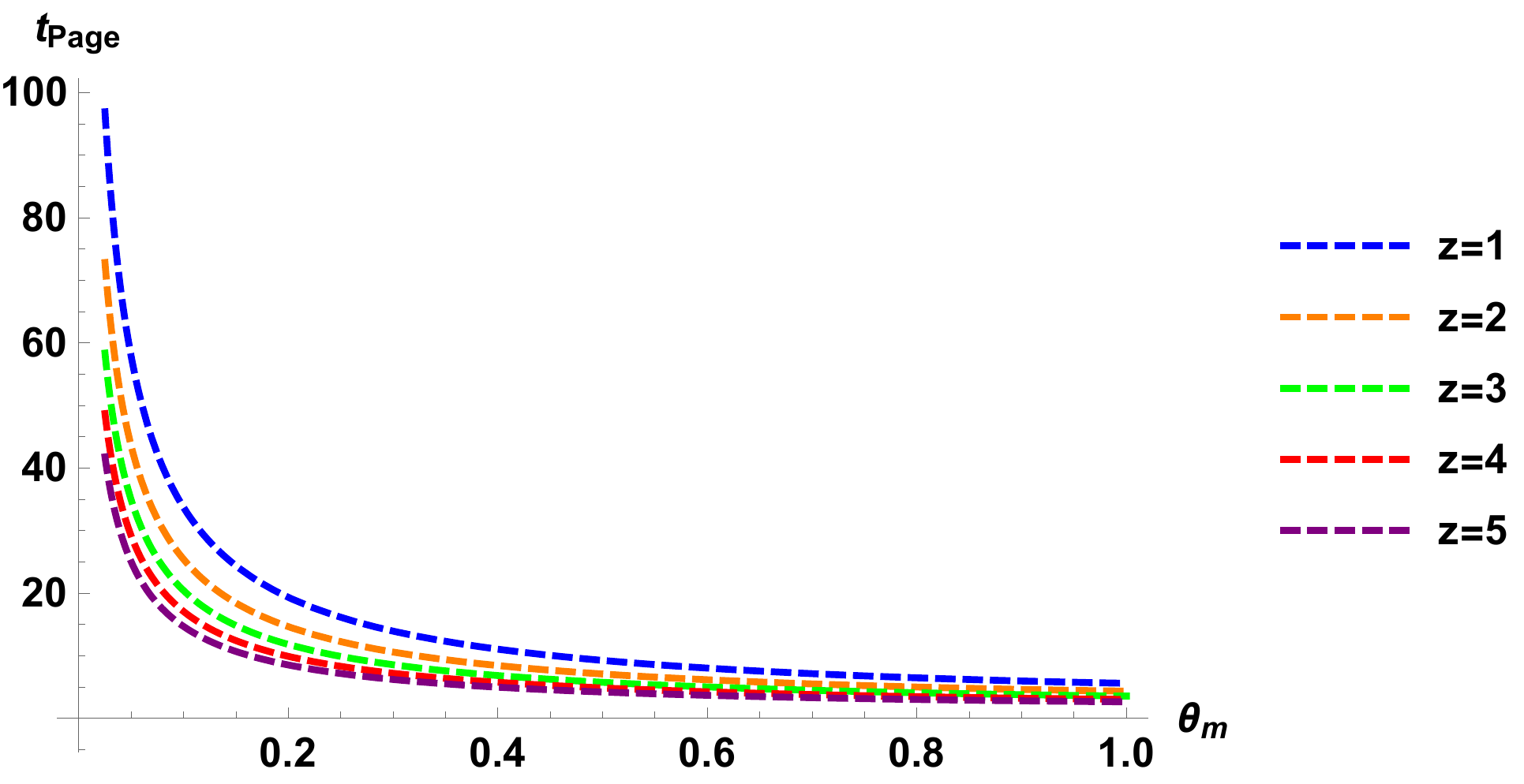}
		\hspace{0.2cm}
		\includegraphics[scale=0.37]{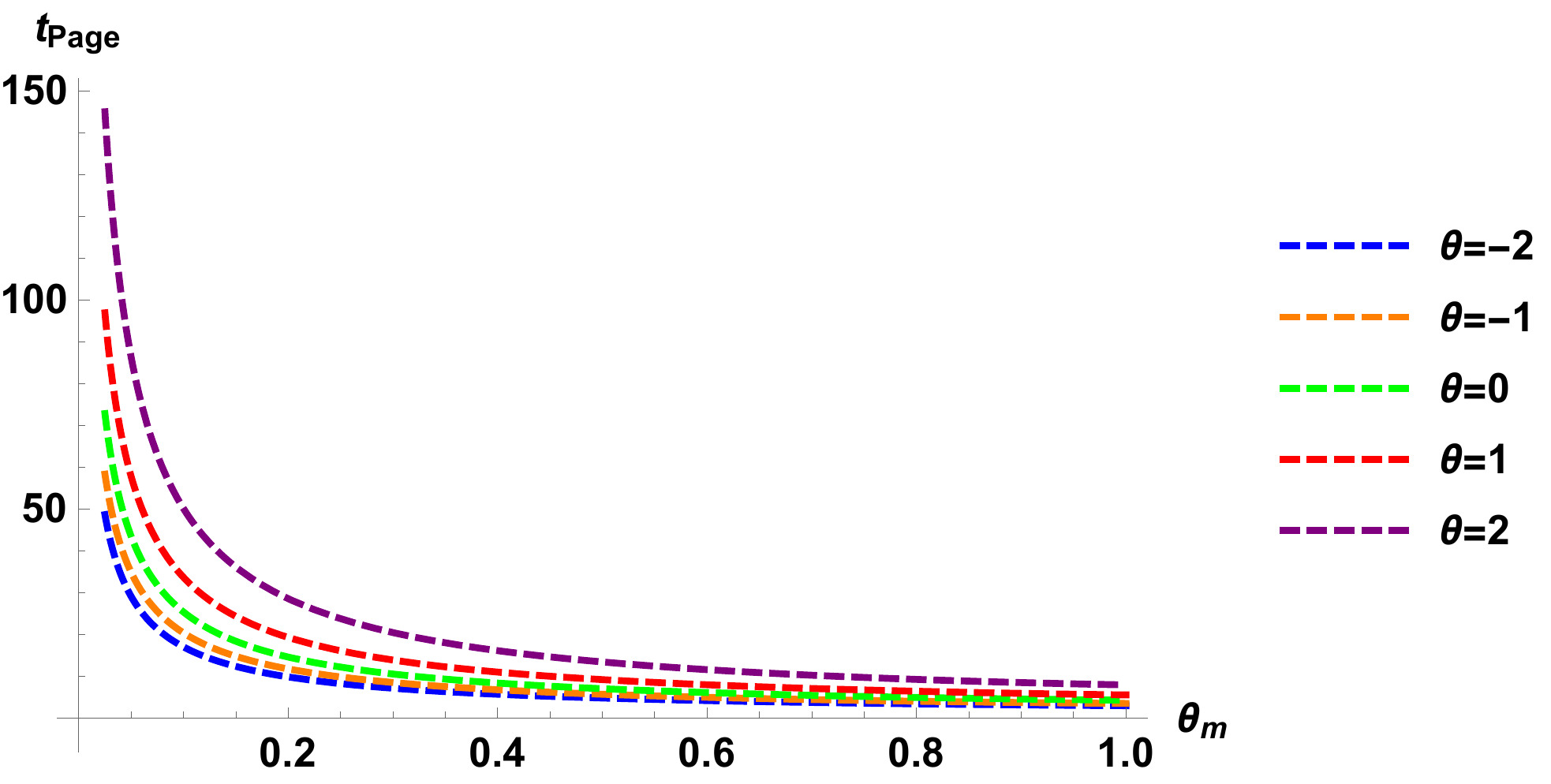}
	\end{center}
	\caption{
		Page time when the matter is described by a HV $QFT_{d+2}$ with $\theta_m \neq 0$ as a function of $\theta_m$: {\it Left}) for $d=2$, $\theta=0$ and different values of $z$. {\it Right}) for $d=3$, $z=1$ and different values of $\theta$. Here we set $A=V_d=r_h=1$ and $G_{N,r} =0.001$.
	}
	\label{fig: t-page-thetam neq 0-thetam}
\end{figure}

As mentioned before, the EE of the matter fields at early times is given by that of the connected RT surfaces (See the left panel of figure \ref{fig: RT-surfaces}). By applying eqs. \eqref{d} and \eqref{EE-de neq 1-2d}, one has
\bea
S_{\rm matter}^{\rm f} 
& = & S_{\rm matter}^{\rm f} (a_-, a_+) + S_{\rm matter}^{\rm f} (b_-, b_+)
\cr && \cr
&=& \frac{A \Upsilon_0^{1- \theta_m}}{3 \theta_m} \Bigg[ \!\! 
\left( \!\!  \left( \!\! \frac{\beta a^{z- \theta_e} \sqrt{f(a)}}{2 \pi}\right)  \!\! \cosh \left( \frac{2 \pi t_a}{\beta}\!\! \right) \!\! \right)^{\!\!  \theta_m} 
\!\!\!\! + \left( \!\! \left( \frac{ \beta}{2 \pi} \right) \cosh \left( \frac{2 \pi t_b}{\beta}\right) \!\! \right)^{ \!\!  \theta_m} \!\! 
\Bigg] \! .
\label{S-matter-islands-theta-neq-0-early-times}
\eea 
After adding the gravitational part, the generalized entropy is given by
\bea
S_{\rm gen}  &=&  \frac{V_d a^{d_e}}{2 G_{N,r}} + \frac{A \Upsilon_0^{1- \theta_m}}{ \!\! 3 \theta_m} \Bigg[
\left( \left( \frac{\beta a^{z- \theta_e} \sqrt{f(a)}}{2 \pi}\right)  \cosh \left( \frac{2 \pi t_a}{\beta}\right) \right)^{ \!\!  \theta_m}
\!\!  
\cr && \cr
&& \;\;\;\;\;\;\;\;\;\;\;\;\;\;\;\;\;\;\;\;\;\;\;\;\;\;\;\;\;\;\;\;\;\;\;\;\;\;\;\;\;\;\; + \left( \left( \frac{ \beta}{2 \pi} \right) \cosh \left( \frac{2 \pi t_b}{\beta}\right) \right)^{\theta_m}
\Bigg]
\cr && \cr
&=&
\frac{V_d a^{d_e}}{2 G_{N,r}} + \frac{A \Upsilon_0^{1- \theta_m}}{3 \theta_m} \Bigg[
\left( \frac{\beta a^{z- \theta_e} \sqrt{f(a)}}{2 \pi}\right)^{ \!\! \theta_m}  \!\! \left( 1 + \frac{2 \pi^2 \theta_m t_a^2}{\beta^2}\right)
\cr && \cr
&& \;\;\;\;\;\;\;\;\;\;\;\;\;\;\;\;\;\;\;\;\;\;\;\;\;\;\;\;\;\;\;\;\;\;\;\;\;\;\;\;\;\;\;
+ \left( \frac{ \beta}{2 \pi} \right)^{ \!\! \theta_m} \!\! \left( 1+ \frac{2 \pi^2 \theta_m t_b^2}{\beta^2}\right)
\Bigg].
\label{S-gen-theta-neq-0-islands-early-times}
\eea 
Next, by extremizing the above expression with respect to $t_a$, one obtains $t_a=0$. On the other hand, after extermization with respect to $a$ one arrives at
\bea
\frac{\partial S_{\rm gen}}{\partial a} = \frac{d_e V_d r_h^{d_e -1}}{2 G_N} + \frac{A  r_h^{ - \theta_m \left( \theta_e + \frac{1}{2} \right)} }{3 (d_e +z)^\frac{\theta_m}{2}} \left( \frac{\Upsilon_0}{2} \right)^{1- \theta_m} (a -r_h)^{\frac{\theta_m}{2} -1} + \mathcal{O} \left( (a- r_h)^{\frac{\theta_m}{2}} \right) \! ,
\eea 
which gives the following expression for $a$
\bea
a = r_h + \left( - \frac{ A G_{N,r} 2^{\theta_m}  \Upsilon_0^{1- \theta_m} }{ 3 V_d d_e (d_e +z)^{\frac{\theta_m}{2}}  r_h^{ \left( d_e - 1+ \theta_m \left(  \theta_e +\frac{1}{2}\right) \right) } }\right)^{\frac{2}{2 - \theta_m}}.
\label{a-theta-neq 0-islands-early-times}
\eea 
It is straightforward to verify that for these values of $t_a$ and $a$, the entropy of Hawking radiation is imaginary. Therefore, one might conclude that there are no islands at early times.
\\On the other hand, at late times, the EE of matter fields is given by that of the disconnected RT surfaces. In this case, by plugging eq. \eqref{EE-de neq 1-2d} into eq. \eqref{S-matter-theta-0-islands-late-times}, one has
\bea
\label{S-matter-islands-theta-neq-0-late-times}
S_{\rm matter}^{\rm f} 
& = & S_{\rm matter}^{\rm f} (a_+, b_+) + S_{\rm matter}^{\rm f} (a_-, b_-) 
\cr && \cr
& = & \frac{A \left( 2 \Upsilon_0 \right)^{1- \theta_m} }{3 \theta_m} d(a_+,b_+)^{\theta_m}
\\
& = & \frac{A \left( 2 \Upsilon_0 \right)^{1- \theta_m} }{3 \theta_m} \Bigg[
\! \frac{ \beta^2 a^{(z - \theta_e)} \sqrt{f(a)}}{2 \pi^2} \left( \!\! \cosh \left( \frac{2 \pi}{ \beta} (b - r^*(a)) \right) \!\! - \! \cosh \left( \frac{2 \pi}{ \beta} (t_a - t_b) \right) \!\!\right)
\!\! \Bigg]^\frac{\theta_m}{2}
\nonumber
\eea 
At late times, by applying the approximation given in eq. \eqref{approx-late-times-2-1}, one obtains
\bea
S_{\rm gen} & = &  \frac{V_d a^{d_e}}{2 G_{N,r}} + \frac{A \left( 2 \Upsilon_0 \right)^{1- \theta_m} }{3 \theta_m}
\left( \frac{\beta^2 a^{(z-\theta_e)} \sqrt{f(a)}}{ 4 \pi^2} e^{\frac{2 \pi}{\beta} (b - r^*(a))} \right)^{\frac{\theta_m}{2}} 
\cr&&\cr
&& 
\;\;\;\;\;\;\;\;\;\;\;\;\;\;\;\; \times \Bigg[ 
1 + \frac{\theta_m}{2} e^{\frac{4 \pi}{\beta} (r^*(a) - b)} 
 - \theta_m e^{\frac{2 \pi}{\beta} ( r^*(a) - b)} \cosh \left( \frac{2 \pi}{\beta} (t_a - t_b) \right)
\Bigg].
\label{S-gen-islands-theta neq 0-late-times}
\eea 
Next, by extremizing $S_{\rm gen}$ with respect to $t_a$, one finds
\bea
t_b = t_a.
\label{ta=tb-theta neq 0}
\eea 
On the other hand, by plugging eq. \eqref{ta=tb-theta neq 0} into eq. \eqref{S-gen-islands-theta neq 0-late-times} and extremizing $S_{\rm gen}$ with respect to $a$, one has
\bea
\frac{ \partial S_{\rm gen}}{ \partial a} = B_1 + \frac{B_2}{\alpha} + \mathcal{O}  (\alpha)=0,
\label{ext-S-gen-islands-theta neq 0}
\eea
where $\alpha = \sqrt{ \frac{a- r_h}{r_h}}$ and
\bea
B_0 & = & \gamma - b \; r_h^z (d_e +z) + \psi \left( \frac{z}{d_e +z}\right),
\cr && \cr
B_1 & = & \frac{V_d r_h^{d_e -1} d_e}{2 G_N} + \frac{A e^{- \frac{\theta_m}{4}  B_0 } \Upsilon_0^{1 - \theta_m}}{6 d \theta_m (d_e +z)^{ \theta_m} r_h^{1 + \frac{\theta_m}{2} ( z + \theta_e)}} 
\Bigg[ 2d (d_e +z) \theta_m  e^{ B_0 }
\cr && \cr
&& \;\;\;\;\;\;\;\;\;\;\;\;\;\;\;\;\;\;\;\;\;\;\;\;\;\;\;\;\;\;\;\;\;\;\;\;\;\;\;\;\;\;\;\;\;\;\;\;\;\;\;\;\;\;\;\;\;\;\;\;\;\;\;\;\; + 2d ( z+ 1- d_e) - \theta_m \left( 2 \theta + d ( d_e - 2z) \right)
\Bigg] ,
\cr&&\cr
B_2 & = & - \frac{A (d_e +z)^{\frac{1}{2} - \theta_m}}{3 r_h^{1+ \frac{ \theta_m}{2} ( z + \theta_e)}}  \Upsilon_0^{1 - \theta_m} e^{ \frac{ (2- \theta_m) B_0}{4}}.
\eea 
From eq. \eqref{ext-S-gen-islands-theta neq 0}, one simply obtains
\bea
a = r_h + \left( \frac{ 2 A G_{N,r}}{3 V_d d_e }  \right)^2 \frac{ (d_e +z)^{1- 2 \theta_m} \; e^{ \frac{( 2 -\theta_m)}{2} \left( \gamma - b \; r_h^z (d_e +z) + \psi \left( \frac{z}{d_e +z}\right)\right)} }{ \Upsilon_0^{2 ( \theta_m -1)} r_h^{2d_e - 1 + \theta_m (z + \theta_e) }}  + \mathcal{O} \left( G_{N,r}^3 \right) \!.
\label{a-theta neq 0}
\eea
At the end, by plugging eqs. \eqref{ta=tb-theta neq 0} and \eqref{a-theta neq 0} into eq. \eqref{S-gen-islands-theta neq 0-late-times}, the entropy of Hawking radiation is obtained as follows
\bea
S_{ \mathcal{R}} = 2 S_{th} + \frac{2 A \; e^{- \frac{\theta_m}{4} \left( \gamma - b (d_e +z) r_h^z + \psi \left( \frac{z}{d_e+z}\right) \right)}}{3 \theta_m (d_e +z)^{\theta_m} r_h^{\frac{1}{2} (z + \theta_e) \theta_m }}  \left( \frac{\Gamma \left( \frac{1}{2(1-\theta_m)}\right)}{\sqrt{\pi} \Gamma \left( \frac{( \theta_m-2)}{2(\theta_m-1)}\right)}\right)^{\theta_m-1},
\label{S-R-theta neq 0-islands-late-times}
\eea 
which is constant in time. Therefore, $S_R$ saturates when there is an island. In figure \ref{fig: S-R-theta neq 0}, the entropy of Hawking radiation is plotted as a function of $z$, $\theta$ and $\theta_m$.  At late times, before the Page time, $S_R$ shows an exponential growth in time (See eq. \eqref{S-R-theta neq 0-no island-late-times}). However, after the Page time, it saturates (See eq. \eqref{S-R-theta neq 0-islands-late-times}).
\\On the other hand, by equating eqs. \eqref{S-R-theta neq 0-no island-late-times} and \eqref{S-R-theta neq 0-islands-late-times}, the Page time is simply obtained as follows
\bea
t_{\rm Page} = \frac{2}{ \theta_m ( d_e +z) r_h^z} \log \left( \frac{3 \theta_m ( d_e +z)^{\theta_m} V_d \; r_h^{ (d_e + z \theta_m)} \Upsilon_0^{\theta_m-1} }{2 A G_{N,r} } \right).
\label{t-page-theta-neq-0}
\eea 
Therefore $t_{\rm page} \propto \log { \left( \frac{1}{G_{N,r}}\right)}$, which is a consequence of the exponential growth of $S_R$ with time before reaching the page time (See eq. \eqref{S-R-theta neq 0-no island-late-times}).
In figure \ref{fig: t-page-thetam neq 0-z-theta}, the Page time is plotted as a function of $z$ and $\theta$ of the black brane. It is observed that $t_{\rm Page}$ is a decreasing function of the exponent $z$. However, it is an increasing function of $\theta$. On the other hand, in figure \ref{fig: t-page-thetam neq 0-thetam}, the Page time is plotted in terms of $\theta_m$ which shows that it is a decreasing function of $\theta_m$.
\section{Discussion}
\label{Sec: Discussion}

In this paper we studied the Page curve for a two-sided Hyperscaling Violating (HV) black brane in $d+2$ dimensions. We assumed that the matter fields in the black brane geometry and inside the baths are the same. Moreover, we considered two different situations for the matter fields: First, they are described by a $CFT_{d+2}$. Second, they are described by a d+2 dimensional HV QFT which is dual to a d+3 dimensional gravity that is a HV geometry at zero temperature (See eq. \eqref{metric-HV-zero-temp}). In both cases,  it was observed that at very early times there are no islands. However at late times the presence of an island is necessary to obtain a correct Page curve for the entropy of the Hawking radiation. 
\\For matter CFT, at early times there are no islands. In this case, the entropy of Hawking radiation $S_{\rm R}$ shows a quadratic growth in time (See eq. \eqref{S-R-CFT-no islands-early-times}). Next, it grows linearly with time (See eq. \eqref{S-R-CFT-no islands-late-times}). If there are no islands, $S_{\rm R}$ exceeds twice the coarse-grained (thermodynamic) entropy $S_{\rm th}$ of the black brane. However, when there is an island, $S_{\rm R}$ becomes a constant (See eq. \eqref{S-R-CFT-island-late-times}). The corresponding Page curve is plotted in figure \ref{fig: S-R-CFT} for different values of the exponents $z$ and $\theta$ of the black brane. 
On the other hand, the Page time is proportional to $\frac{S_{\rm th} T}{c}$ (See eq. \eqref{t-page-CFT}), where $S_{\rm th}$ and $T$ are the thermal entropy and temperature of the black brane, and $c$ is the central charge of the matter CFT. Moreover, in figure \ref{fig: t-Page-CFT}, the Page time is plotted as a function of $z$ and $\theta$. It was observed that $t_{\rm Page}$ is a decreasing function of $z$. On the other hand, it is an increasing function of $\theta$. Moreover, for $\theta \leq 0$, the Page time is always smaller than that for the case $z=1$ and $\theta=0$. In other words, the entropy of Hawking radiation for HV black branes saturates sooner than that for planar AdS-Schwarzschild black holes, if one has $\theta \leq 0$. However, for positive values of $\theta$, the Page time becomes larger than that for the case $z=1$ and  $\theta=0$, if one decreases $z$.
\\It should be emphasized that for $z=1$ and $\theta =0$, the black brane geometry reduces to a $d+2$ dimensional planar AdS-Schwarzschild black hole. Therefore all of our results can be applied for this type of black hole, if one sets $z=1$ and $\theta=0$. We verified that for $d=2$, our results are consistent with those for a four dimensional planar AdS-Schwarzschild black hole in the critical gravity model reported  in ref. \cite{Alishahiha:2020qza}, if one sets all of the higher derivative couplings in the action to zero.
\\On the other hand, we studied the case where the matter is described by a d+2 dimensional HV QFT. In this case one can assign two exponents $z_m$ and $\theta_m$ to the matter. Moreover, since the EE of matter is independent of the exponent $z_m$, the entropy of Hawking radiation is also independent of $z_m$. In other words, it only depends on the exponent $\theta_m$. We examined the two cases $\theta_m =0$ and $\theta_m \neq 0$ separately. Moreover, since the HV QFT has a dual gravity, we applied the holographic prescription of ref. \cite{Headrick:2010zt} to calculate the EE of the matter fields on the two disjoint intervals $[b_-,a_-] \cup [a_+,b_+]$ when there is an island $\mathcal{I}$ (See figure \ref{fig: RT-surfaces}). In other words, at early times when the two intervals $[b_-,a_-]$ and $[a_+,b_+]$ are close to each other, we considered the connected RT surfaces. However, at late times when the two intervals are very far from each other, we applied the disconnected RT surfaces. It was observed that:
\\For $\theta_m =0$, the behaviors of the Page curve and Page time are the same as those for matter CFT. In other words, at early times there are no islands. In this case, $S_{\rm R}$ shows a quadratic growth in time (See eq. \eqref{S-R-theta-0-no islands-early-times}). Next, it grows linearly with time (See eq. \eqref{S-R-theta-0-no islands-late-times}). If there are no islands, $S_{\rm R}$ again exceeds twice the coarse-grained entropy of the black brane. However, when there is an island, $S_{\rm R}$ becomes a constant (See eq. \eqref{S-R-theta neq 0-islands-late-times}). Moreover, the Page time is proportional to $\frac{S_{\rm th} T}{A}$ (See eq. \eqref{t-page-theta-0}) where $A= \frac{3 R}{2 G_{N}}$.
It should be pointed out that the EE of matter fields, and hence the entropy of Hawking radiation is independent of $z_m$. On the other hand, for $z_m=1$ and $\theta_m=0$, the HV $QFT_{d+2}$ becomes a $CFT_{d+2}$. Therefore, all of the results for $\theta_m=0$ should be the same as those for the case where the matter fields are described by a $CFT_{d+2}$. In particular, the plots of the Page curve and Page time are again given by figures \ref{fig: S-R-CFT} and \ref{fig: t-Page-CFT}, if one replaces $A$ with $c$ in the formulas. This observation shows that by applying the  holographic prescription of ref. \cite{Headrick:2010zt}, the entropy of Hawking radiation obeys the expected Page curve. Furthermore, $t_{\rm Page}$ is a decreasing function of the exponent $z$ and an increasing function of $\theta$ (See figure \ref{fig: t-Page-CFT}). 
\\For $\theta_m \neq 0$, it was verified that at early times there are no islands. In the absence of an island, at the beginning $S_{\rm R}$ again shows a quadratic behavior with time (See eq. \eqref{S-R-theta neq 0-no island-early-times}). Next, it grows exponentially with time (see eq. \eqref{S-R-theta neq 0-no island-late-times}), which is in contrast to the usual linear growth for flat and AdS black holes with matter CFT. If there are no islands, again the entropy of Hawking radiation exceeds twice the coarse-grained entropy of the black brane.
However, when there is an island, $S_{\rm R}$ becomes independent of time (See eq. \eqref{S-R-theta neq 0-islands-late-times}). This behavior is a consequence of the exponential growth of the entropy of Hawking radiation before the Page time. Moreover, the corresponding Page curve is plotted in figure \ref{fig: S-R-theta neq 0} for different values of the exponent $z$, $\theta$ and $\theta_m$. It was also observed that the Page time is proportional to $\log \left( \frac{1}{G_{N,r}} \right)$ (See eq. \eqref{t-page-theta-neq-0}). Furthermore, similar to the case $\theta_m = 0$, the Page time is a decreasing function of  $z$ and an increasing function of $\theta$ (See figure \ref{fig: t-page-thetam neq 0-z-theta}). Moreover, $t_{\rm Page}$ is independent of $z_m$ and is a decreasing function of $\theta_m$ (See figure \ref{fig: t-page-thetam neq 0-thetam}).
\\As mentioned before, for $\theta_m \neq 0$, the EE of radiation grows exponentially before the Page time (see eq. \eqref{S-R-theta neq 0-no island-late-times}). As long as we know, this rate of growth for the EE is surprising. 
\footnote{We would like to thank the referee for her/his enlightening comments on this point.}
More precisely, the rate of growth of the EE for Vaidya black branes with Hyperscaling Violation when the entangling region is in the shape of a strip with width $l$ and lengths $L$ were explored in refs. \cite{Alishahiha:2014cwa,Fonda:2014ula}. It was observed that there are three phases for the growth of the EE: First, early times, i.e. $t \ll  \rho_h^z  \propto \frac{1}{T}$ where $\rho_h$ and $T$ are the horizon radius and temperature of the black brane, it obeys a power law behavior  and is given by
\bea
\Delta S= \frac{L^{d-1} m}{8 G_N (z+1)} (z t)^{ 1 + \frac{1}{z}},
\label{EE-growth-early-time} 
\eea 
where $\Delta S$ is the difference of the EE with that of the vacuum and $m$ is related to the mass of the black brane. Notice that it is independent of $\theta$.
For $z=1$ and $\theta=0$, where the HV QFT becomes a CFT, it is quadratic in time as it is expected for holographic CFTs \cite{Liu:2013iza,Liu:2013qca}. Moreover, for very large $z$, it becomes linear in time.
Second, at intermediate times, i.e. $ \rho_h^z \lesssim t \lesssim \frac{l}{2} \rho_h^{z-1} $, it grows linearly in time
\bea
\Delta S= \frac{L^{d-1}}{2 G_N \rho_h^{d_e +z -1}} v_E t, 
\label{EE-growth-intermediate-time}
\eea 
where $v_E$ is the entanglement velocity and given by
\bea 
v_E= \frac{(\eta-1)^{\frac{\eta-1}{2}}}{ \eta^{\frac{\eta}{2}}},  \;\;\;\;\;\;\;\;\;\;\;\;\;\;\;\;\;\;\; \eta = \frac{2 (d_e + z -1)}{d_e + z}.
\label{v-E-HV}
\eea 
Third, it saturates to a constant value at late times. Therefore, this exponential growth is a new feature and it would be very interesting to investigate it further and to explore whether or not this growth rate is consistent with unitarity.
\\At the end, it would also be interesting to do these calculations for one-sided HV black branes and two-sided charged HV black branes and study the effects of the exponents $z$ and $\theta$ on the Page curve and Page time. 

\section*{Acknowledgment}

We would like to thank Mohsen Alishahiha very much for his support and illuminating comments during this work. We are also very grateful to Mukund Rangamani, Edgar Shaghoulian, Ali Naseh, Amir Hossein Tajdini, Pablo Bueno and Amin Faraji Astaneh for having very helpful discussions.
The work of the author is supported by the school of physics at IPM and Iran Science Elites Federation (ISEF). 




\end{document}